\documentclass[reprint,superscriptaddress, amsmath,amssymb, aps, prb,floatfix]{revtex4-1}

\usepackage{graphicx}
\usepackage{dcolumn}
\usepackage{bm}
\usepackage{hyperref}
\usepackage{xcolor}
\usepackage{lipsum}
\usepackage{transparent}

\begin{document}
\bibliographystyle{apsrev4-1}
\title{Dynamical Charge Susceptibility in the Hubbard Model}
\author{Xinyang Dong}
\affiliation{Department of Physics, University of Michigan, Ann Arbor, MI 48109, USA}
\author{Xi Chen}
\affiliation{Department of Physics, University of Michigan, Ann Arbor, MI 48109, USA}
\affiliation{Center for Computational Quantum Physics, The Flatiron Institute, New York, New York, 10010, USA}
\author{Emanuel Gull}%
\affiliation{Department of Physics, University of Michigan, Ann Arbor, MI 48109, USA}
\affiliation{Center for Computational Quantum Physics, The Flatiron Institute, New York, New York, 10010, USA}

\date{\today}

\begin{abstract}
We compute the dynamical charge susceptibility in the two-dimensional Hubbard model within the dynamical cluster approximation. In order to understand the connection between charge susceptibility and pseudogap, we investigate the momentum, doping, and temperature dependence. We find that as a function of frequency, the dynamical charge susceptibility is well represented by a single peak at a characteristic frequency. It shows little momentum or temperature dependence, while the doping dependence is more evident, and no clear signature of the pseudogap is observed. Data for the doping evolution of the static susceptibility and for fluctuation diagnostics are presented. Our susceptibilities should be directly measurable in future Momentum-resolved electron energy-loss spectroscopy experiments.
\end{abstract}

\maketitle

Charge fluctuations in the high-temperature superconductors have generated renewed interest, as new experimental probes, such as Resonant inelastic X-ray scattering RIXS \cite{Ishii17,Suzuki18,Neto18,Hepting18,Miao19,Ishii19} or Momentum-resolved electron energy-loss spectroscopy M-EELS \cite{Sean17,Hussian19} may be able to directly measure them as a function of momentum and energy.

Much of the low energy physics of the cuprates is believed to be described by the one-band Hubbard model with an interaction strength close to the bandwidth and a small next-nearest neighbor hopping \cite{Anderson87, Scalapino07}. The model has a Mott-insulating state at half filling, as well as pseudogap, superconducting, and Fermi liquid phases in regimes that are remarkably similar to what is seen in the cuprates. 
Hubbard model simulations of spectral function \cite{Huscroft01,Civelli05,Stanescu06,Parker08,Liebsch09,Ferrero09,Ferrero09B,Sakai09,Sakai10,Lin10} show a clear suppression of the density of states at the antinodal (but not at the nodal) points in the pseudogap regime. Results for optical conductivity \cite{Ioffe00,Toschi05,Millis05,Comanac08,Ferrero10,Lin10,Bergeron11}, Raman spectroscopy \cite{Lin12,Gull13_raman}, neutron spectroscopy \cite{LeBlanc19_neutron} and Nuclear magnetic resonance \cite{Macridin06,Chen17} similarly exhibit the salient features of experiments.

Much less is known about the charge susceptibility of the Hubbard model. Ground state calculations with a variety of methods show various types of charge order at $1/8$ doping \cite{White03_stripe,Hager05_stripe,Chang08_strip, Scalapino12_stripe,Zheng16_stripe,Huang17_stripe,Zheng17_stripe,Huang18_stripe,Vanhala18_stripe,Darmawan18_stripe,Tocchio19_stripe}, competing closely with superconductivity. High-temperature lattice simulations show the doping evolution of the dynamical susceptibility far above pseudogap regime \cite{Zimmermann97,Sherman08,Hochkeppel08,Devereaux15}. However, no theoretical results are available so far in the pseudogap regime or near the superconducting phase. These results are needed for the attribution of experimental results to pseudogap physics, as experiments often measure a combination of low-energy physics and higher energy contributions from three-dimensional and atomic physics \cite{Hepting18,Lin19}.

In this paper, we present a detailed simulation of the momentum, doping, and temperature dependence of the dynamical charge susceptibility $\chi_{ch}(Q, \Omega)$ using eight-site cluster dynamical mean field theory. We also analyze the change of $\chi_{ch}(Q, \Omega)$ in the pseudogap and show the extent to which the establishment of a pseudogap can be understood as a consequence of charge fluctuations.

\begin{figure}[tb]
    \centering
    \includegraphics[width=0.95\columnwidth]{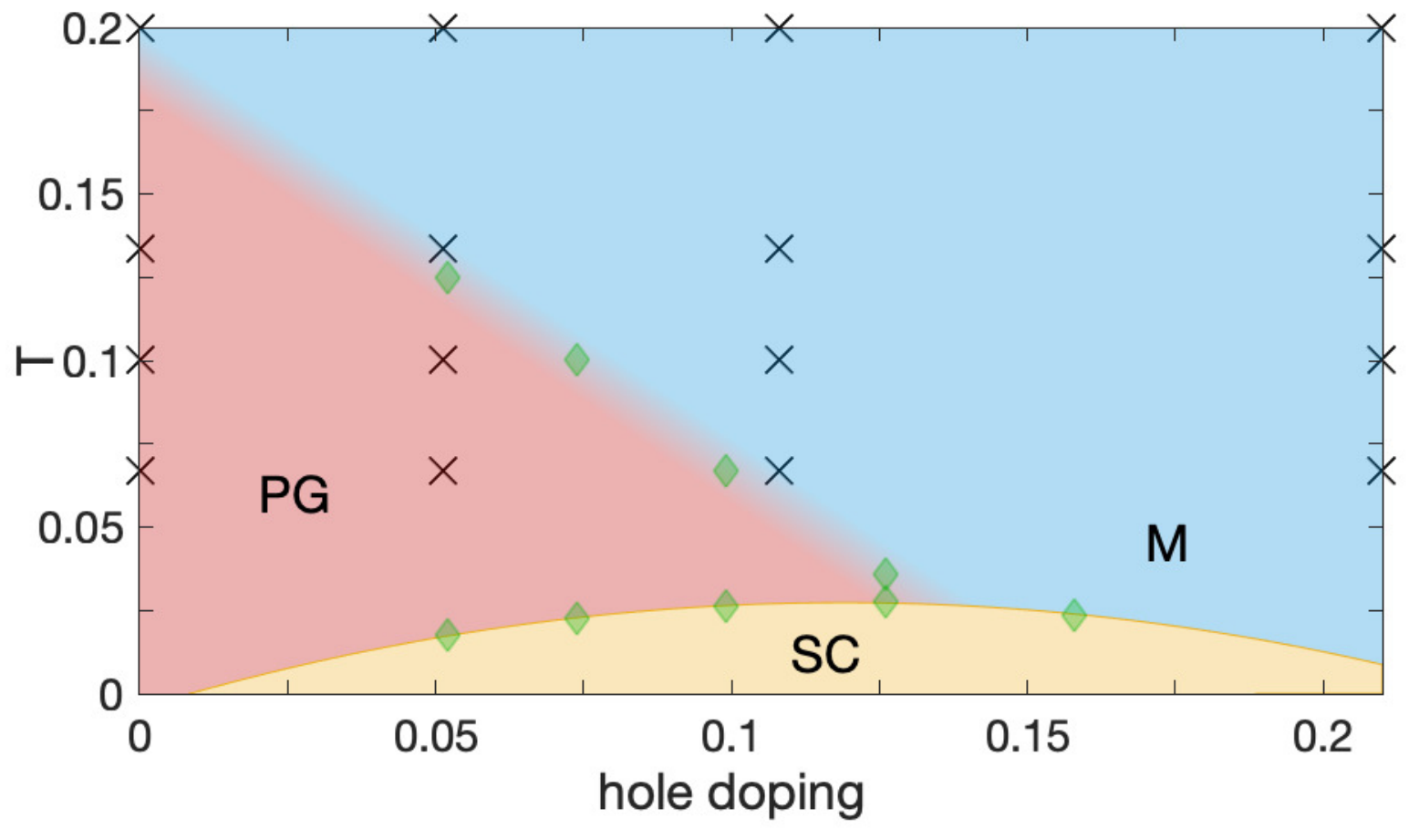}
    \llap{\put(-80,60)
     {\transparent{0.8}\includegraphics[height=2.2cm]{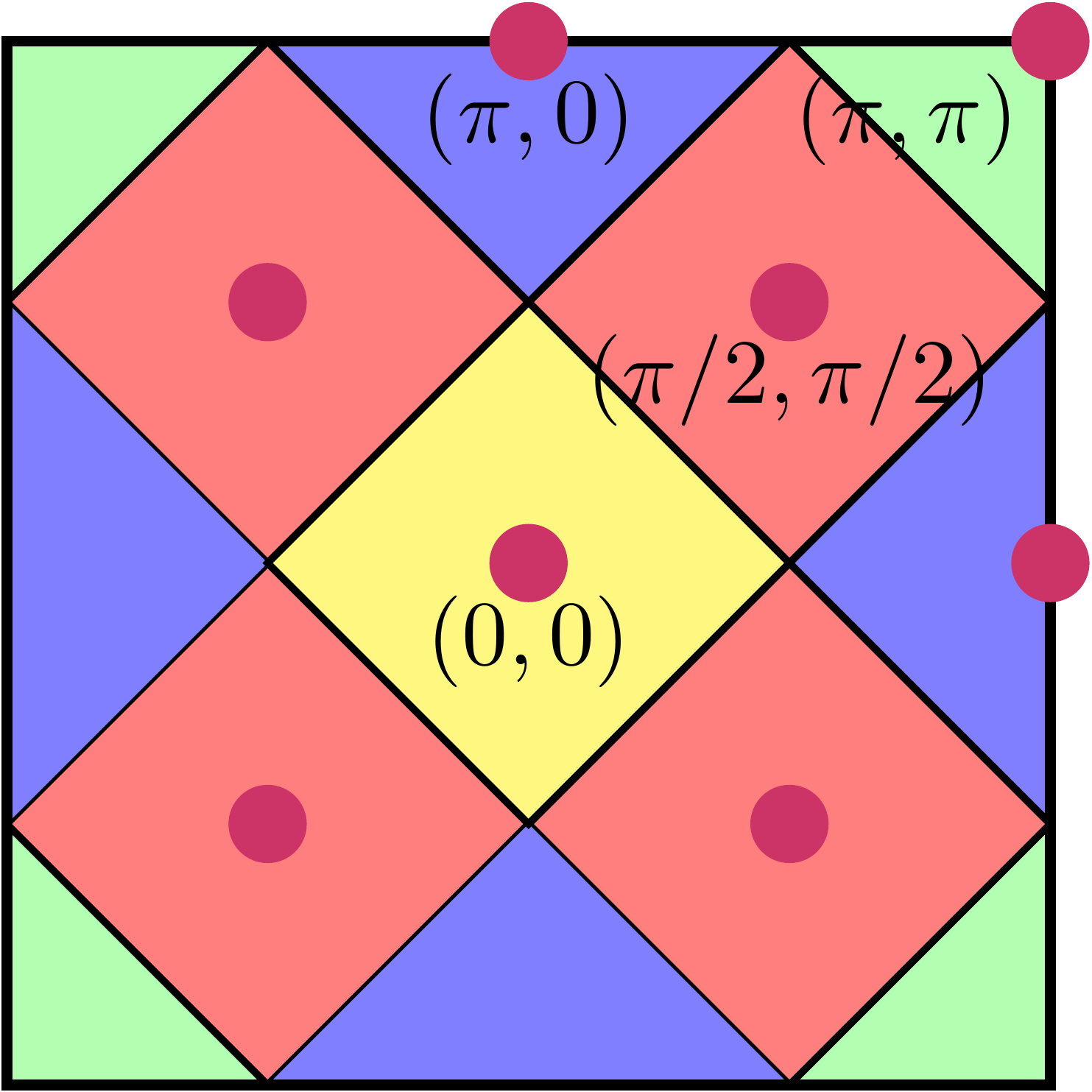}}
    }
    \caption{eight-site DCA phase diagram of the Hubbard model, with metal (M, blue), pseudogap (PG, red), and superconducting (SC, yellow) regions. Crosses denote data points shown in this paper, diamonds points used to extract the phase boundaries. In the paper we use $\delta < 0$ for hole doping. Inset: geometry of the eight-site DCA cluster.}
    \label{fig:PDOverview} 
\end{figure}

\begin{figure*}[bt]
    \centering
    \includegraphics[width=0.95\textwidth]{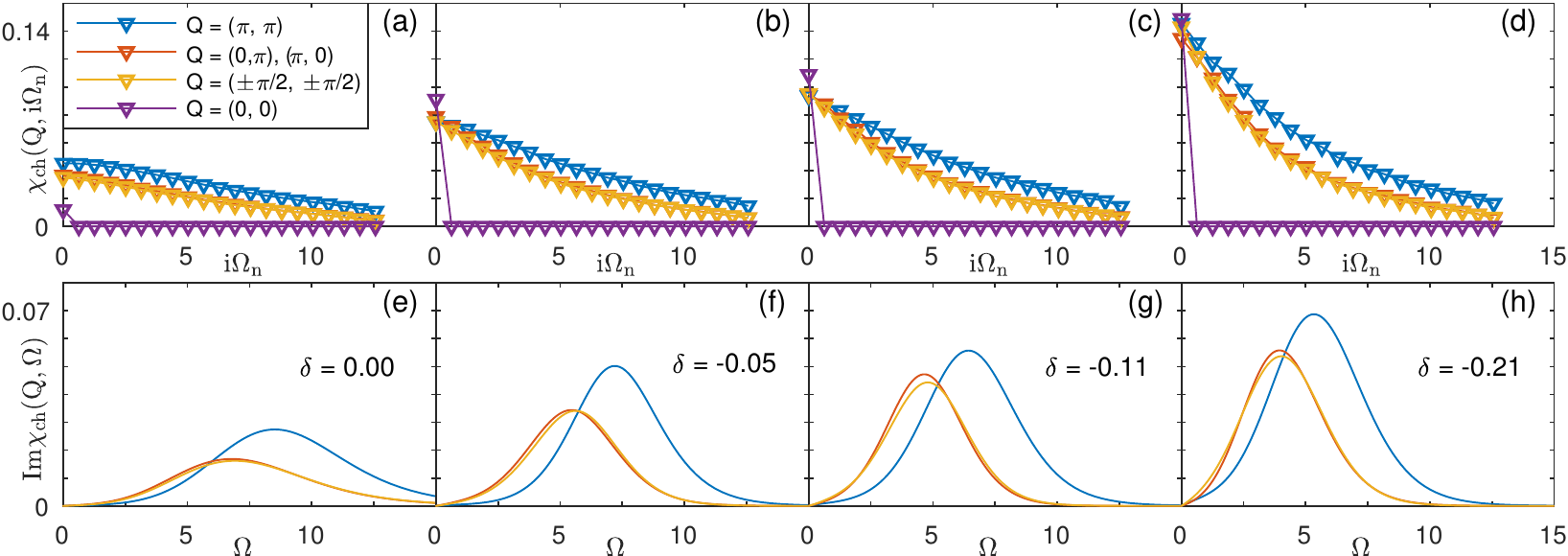}
    \caption{Charge susceptibility at $\beta = 10$ with momentum transfers $Q$ at [(a) and (e)] $\delta = 0$, [(b) and (f)] $\delta = -0.05$, [(c) and (g)] $\delta = -0.11$, and [(d) and (h)] $\delta = -0.21$. [(a)-(d)] are results in Matsubara frequencies, [(e)-(h)] are analytically continued results in real frequency.}
    \label{doping_diffQ}
\end{figure*}

We solve the two-dimensional Hubbard model with repulsive interaction on a square lattice,
\begin{equation}
H=\sum_{k\sigma}(\varepsilon_k-\mu)c^\dagger_{k\sigma}c_{k\sigma}+U\sum_in_{i\uparrow}n_{i\downarrow}
\label{H},
\end{equation}
within the dynamical cluster approximation (DCA) \cite{Maier05} on an eight-site cluster. 
Here $i$ labels the sites in a lattice and $k$ a momentum in the corresponding Brillouin zone. $\varepsilon_k=-2t(\cos k_x + \cos k_y) -4t' \cos k_x \cos k_y$ denotes the dispersion, $\mu$ the chemical potential, and $U$ the on-site interaction. We chose $U/t=7$ and $t'/t=-0.15$ to differentiate electron from hole doping. Where we present data in K or eV, we use t $\approx$ 0.35~eV corresponding to high-$T_c$ superconductors \cite{Anderson94,Anderson95,Kim98_t,Hasan00}. 
At these parameters, the model is Mott insulating at half filling, and superconducting at low temperature (the maximum $T_c$ on the eight-site cluster is near $\mathrm{T}=1/40~t$ \cite{Gull09_8site,Gull12_energy,Gull13_super,Gull14_pairing,Gull15_qp}), and exhibits a pseudogap region where the single-particle spectral function at the node stays metallic while the antinode is insulating \cite{Gull09_8site,Gull10_clustercompare}.
Fig.~\ref{fig:PDOverview} shows an overview of this phase diagram.
The DCA can be understood as a momentum-space approximation of the self-energy that coarse-grains the momentum structure, but retains the full frequency dependence \cite{Maier05,Fuhrmann07}. The method is controlled, in the sense that for cluster size $N_c \rightarrow \infty$ the exact solution is recovered \cite{Maier05_dwave,Fuchs11,LeBlanc15}. Our solution is restricted to the paramagnetic phase.
Inset of Fig.~\ref{fig:PDOverview} shows the cluster geometry.

The main result of this paper is the dynamical charge susceptibility $\chi_{ch}(\mathrm{Q}, \Omega)$ obtained by solving the DCA equations using a numerically exact continuous-time auxiliary field quantum impurity solver \cite{Gull08,Gull11,Gull10_submatrix}. The single and two-particle Green's functions $G_\sigma(k_1\tau_1,k_2\tau_2) $ $=$ $ \langle T_\tau (c_{k_1\sigma}^\dagger (\tau_1) c_{k_2\sigma}(\tau_2))\rangle$ and $G_{2, \sigma_1\sigma_2\sigma_3\sigma_4}(k_1\tau_1,...,k_4\tau_4) $ $=$ $ \langle T_\tau (c_{k_1\sigma_1}^\dagger (\tau_1) c_{k_2\sigma_2}(\tau_2) c_{k_3\sigma_3}^\dagger (\tau_3) c_{k_4\sigma_4}(\tau_4))\rangle$ define the generalized susceptibility as
$\chi_{\sigma \sigma'}(k_1\tau_1,k_2\tau_2,k_3\tau_3)$ $=$ $G_{2,\sigma\sigma\sigma'\sigma'}(k_1\tau_1,k_2\tau_2,k_3\tau_3,0) - G_{\sigma}(k_1\tau_1,k_2\tau_2)G_{\sigma'}(k_3\tau_3,0)$, or  in Fourier space as \cite{Roheringer12} 
\begin{equation}
\begin{aligned}
\chi_{\sigma \sigma'}^{\omega\omega'\Omega}&(k,k',q) = \int_{0}^{\beta} e^{-i\omega\tau_1}e^{i(\omega+\Omega)\tau_2}e^{-i(\omega'+\Omega)\tau_3}\\&\times \chi_{\sigma\sigma'}(k\tau_1,(k+q)\tau_2,(k'+q)\tau_3) d\tau_1 d\tau_2 d\tau_3. \label{eq:chiph}
\end{aligned}
\end{equation}
where $\omega$, $\omega'$ are fermionic and $\Omega$ bosonic Matsubara frequencies. 

\begin{figure}
    \centering
    \includegraphics[width=0.95\columnwidth]{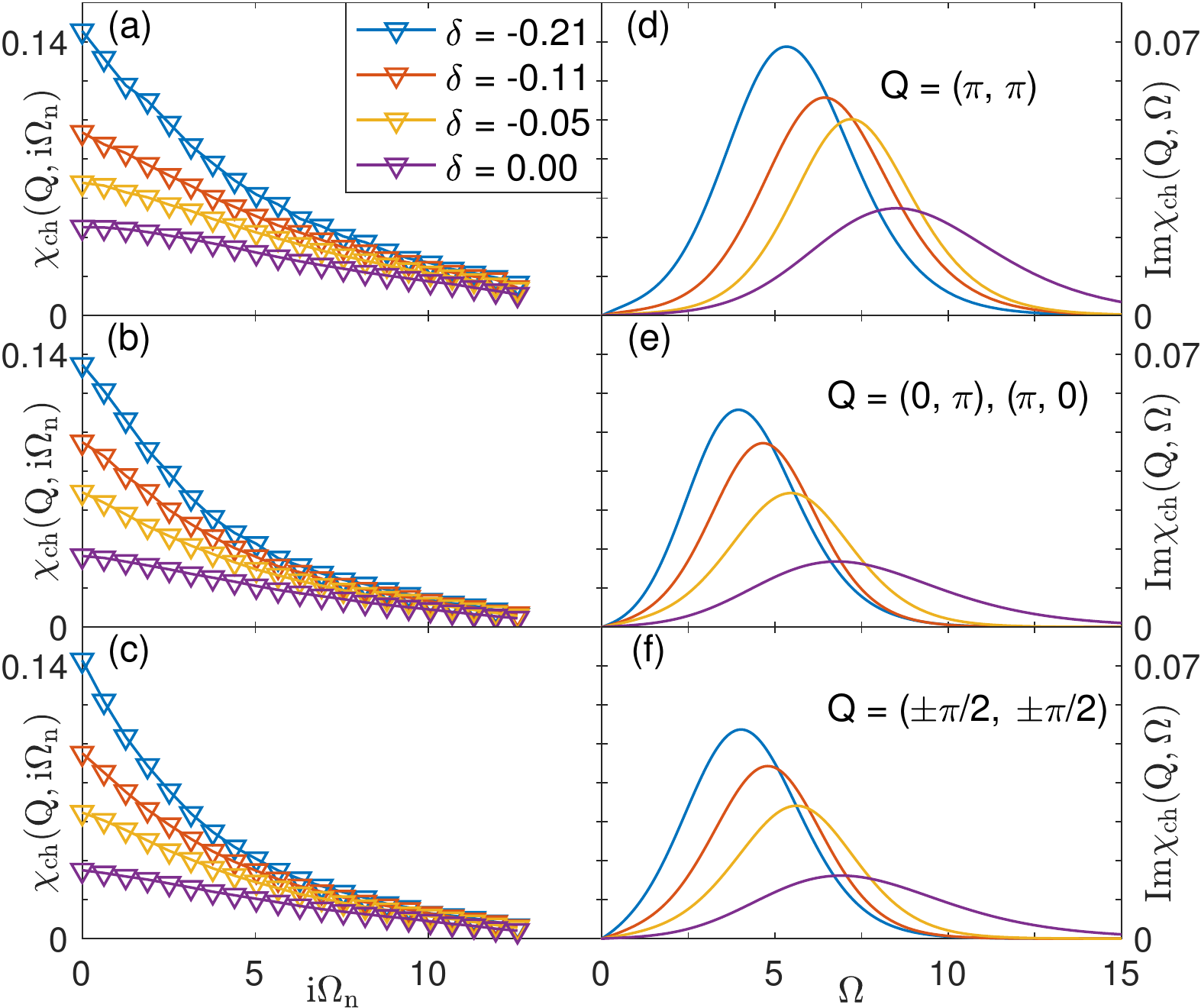}
    \caption{Charge susceptibility at $\beta = 10$ and different dopings with momentum transfer [(a) and (d)] $Q = (\pi, \pi)$, [(b) and (e)] $Q = (0, \pi), (\pi, 0)$, [(c) and (f)] $Q = (\pm \pi/2, \pm \pi/2)$. [(a)-(c)] Results in Matsubara space, [(d)-(f)] results in real space.}
    \label{doping_diffmu}
\end{figure}

The susceptibility in the density channel is defined as 
\begin{align}
    \chi_{d}^{\omega\omega'\Omega}(k, k', q)=\chi_{\uparrow\uparrow}^{\omega\omega'\Omega}(k, k', q) +\chi_{\uparrow\downarrow}^{\omega\omega'\Omega}(k, k', q),
\end{align}\\
and can be decomposed into two parts:
\begin{align}
    \chi_{d}^{\omega\omega'\Omega}&(k, k', q) = \chi_0^{\omega\omega'\Omega}(k, k', q) - \frac{1}{\beta^2 N^2}\chi_0^{\omega\omega_1\Omega}(k, k_1, q)\nonumber \\
    &\times F_{d}^{\omega_1\omega_2\Omega}(k_1, k_2, q)\chi_{0}^{\omega_2\omega'\Omega}(k_2, k', q),
    \label{eq:chid}
\end{align}
with $F_d$ the full vertex in density channel, $N$ the number of momentum points summed up, and $\chi_0$ the bare susceptibility 
$\chi_0^{\omega\omega'\Omega}(k, k', q) = -\beta N G_\sigma (i\omega, k)G_\sigma (i\omega+i\Omega, k+q)\delta_{\omega\omega'}\delta_{k k'}$.

A DCA calculation only yields the ``cluster'' Green's functions and susceptibilities $\chi_{c,d}^{\omega\omega'\Omega}(K, K', Q)$ at the cluster momenta  $K, K'$ and $Q$. The corresponding lattice susceptibility $\chi_{l,d}(k, k', Q)$ is related to this quantity as $\chi_{l,d}^{-1}(k, k', Q) - \chi_{l,0}^{-1}(k, k', Q) = \chi_{c,d}^{-1}(K, K', Q) - \chi_{c,0}^{-1}(K, K', Q)$ (for details see Ref.~\onlinecite{Maier05}). The dynamical charge susceptibility is then obtained as
\begin{align}
\chi_{ch}(Q, \Omega) = \frac{2}{\beta^2 N^2}\sum_{k,k'}\sum_{\omega,\omega'}\chi_{l,d}^{\omega\omega'\Omega}(k, k', Q).
\label{eq:chic}
\end{align}
Eight-site DCA yields results for eight $Q$; four of those are equivalent because of symmetry (see inset of Fig.~\ref{fig:PDOverview}), such that we have four independent momenta: $Q = (0, 0)$, $Q = (\pi/2, \pi/2)$, $Q = (\pi, 0)$, and $Q = (\pi, \pi)$. We have also performed simulations for 4-site and 16-site clusters at select points and high temperature; these results show qualitatively the same results as 8-site clusters.

\begin{figure*}[bth]
    \centering
    \includegraphics[width=0.95\textwidth]{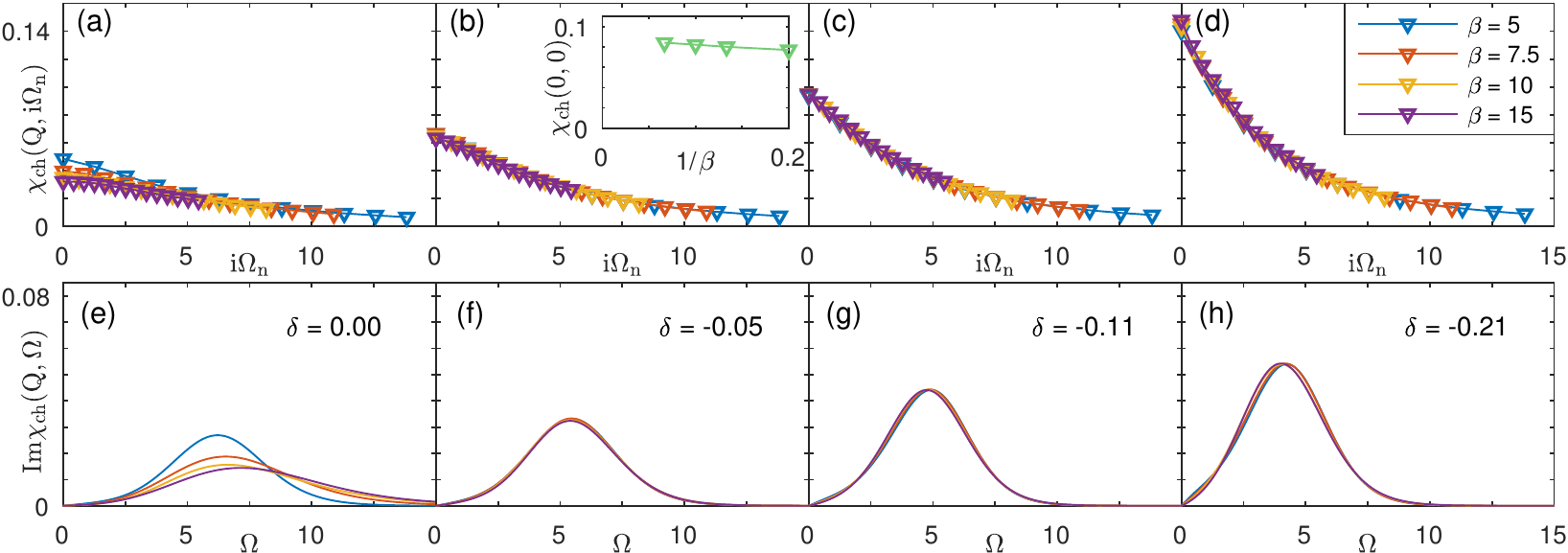}
    \caption{Charge susceptibility at different temperatures with momentum transfer $Q = (\pm \pi/2, \pm \pi/2)$ at doping level [(a) and (e)] $\delta = 0$, [(b) and (f)] $\delta = -0.05$, [(c) and (g)] $\delta = -0.11$, [(d) and (h)] $\delta = -0.21$. [(a)-(d)] Results in Matsubara space, [(e)-(h)] results in real space. Inset of (b): Static uniform charge susceptibility ($Q = (0, 0), \Omega = 0$) at doping $\delta = -0.05$.}
    \label{doping_difftemp}
\end{figure*}  

We present our simulation data as a function of Matsubara frequency $i\Omega_n$. In addition, we also show analytically continued result of $\mathrm{Im}\chi_{ch}(\Omega)$ as a function of real frequency $\Omega$. Analytical continuation~\cite{Jarrell96, Ryan17} is an uncontrolled procedure that may exponentially amplify statistical uncertainties, especially in the case of bosonic functions such as the charge susceptibility considered here.
Values of $\mathrm{Im}\chi(\Omega)$ at high frequencies are generally less reliable than at low frequency. Our data mostly results in a single large peak at a characteristic frequency, and continuations with different default models do not lead to appreciable differences in this feature. As we will show below, a simple interpretation of this feature in terms of single-particle quantities is not possible due to the importance of vertex functions.

Figure.~\ref{doping_diffQ} shows $\chi_{ch}(Q)$ as a function of frequency at T = 0.1t $\sim$ 400~K, at half filling (Figure.~\ref{doping_diffQ} a), and in the underdoped (Figure.~\ref{doping_diffQ} b), optimally doped (Figure.~\ref{doping_diffQ} c), and overdoped (Figure.~\ref{doping_diffQ} d) regime. 
Purple points denote values at $Q = (0, 0)$; the value at $i\Omega_n = 0$ corresponds to the static ($\Omega = 0$) uniform ($Q = (0, 0)$) charge susceptibility which is small in the insulator and generally rises as doping is increased.
A Ward identity \cite{Ward_Identity} requires the frequency dependence to be identically zero in systems that conserve total charge. Unlike many low-order diagrammatic methods \cite{Bohm51,Hedin64,Bickers2004,Brener08,LeBlanc19_neutron}, DCA satisfies this constraint exactly.

Data for susceptibilities at the three-momentum transfer $(\pi, \pi)$ (blue), $(\pi, 0)$ [red, degenerate with $(0, \pi)$], and $(\pi/{2}, \pi/{2})$ [orange, degenerate with $(\pm\pi/{2}, \pm\pi/{2})$] exhibit a smooth frequency dependence.
Remarkably, data at $Q = (\pi/{2}, \pi/{2})$ and at $Q = (\pi, 0)$ are almost identical, both in the half-filled and in the doped case.
The momentum dependence of the charge susceptibility is therefore very different from that of the magnetic susceptibility. For magnetic susceptibility, as found in several approaches \cite{LeBlanc19_neutron, Hochkeppel08, Devereaux15}, the value at $Q = (\pi, \pi)$ is much larger than any other momentum transfer and rapidly grows as temperature decreases. In contrast, within the momentum resolution achievable within DCA, no dominant contribution to the charge susceptibility is found.

In order to make a connection to experiment, we show analytically continued data corresponding to the Matsubara curves in the lower panel [omitting $Q = (0, 0)$]. Within our resolution, our data for $Q = (\pi/{2}, \pi/{2})$ and  $Q = (\pi, 0)$ are described well by a single peak with a maximum near $\omega$ = 7t $\sim$ 2.45~eV (half filling) and 4t $\sim$ 1.40~eV (overdoped). Data at $Q = (\pi, \pi)$ exhibit a peak at a substantially higher frequency (8.5t $\sim$ 2.97~eV for half filling and 5.5t $\sim$ 1.92~eV for overdoped).
 
Figure. \ref{doping_diffmu} shows the doping dependence of our data at constant temperature T = 0.1t $\sim$ 400~K. The top panels show $Q=(\pi,\pi)$, the middle panels $Q=(0,\pi)$, and the bottom panels $Q=(\pi/2,\pi/2)$. Left panels show Matsubara data, right panels the corresponding analytically continued real frequency data.
Four doping points are shown: half filling (purple), underdoped (5\% doping, orange), optimally doped (11\% doping, red), and overdoped (21\% doping, blue).

In Matsubara space, a gradual doping evolution is visible at low frequencies. Zero-frequency values are reduced in comparison to the overdoped values by a factor of about three, while the high-frequency limit remains unchanged. In the real frequency domain, this corresponds to a lowering of the frequency of the charge susceptibility peak, and a general sharpening. As seen previously, no significant momentum dependence is observed, apart from the $\Gamma$ point, which is zero due to charge conservation. This doping evolution is similar to what is found at high temperature \cite{Devereaux15}. 

Figure. \ref{doping_difftemp} shows the temperature dependence of $\chi_{ch}(Q)$ at different doping levels with momentum transfer $Q = (\pm \pi/2, \pm \pi/2)$ at half filling (Figure. \ref{doping_difftemp} a), and in the under-doped (Figure. \ref{doping_difftemp} b), optimally doped (Figure. \ref{doping_difftemp} c), and over-doped (Figure. \ref{doping_difftemp} d) regime. Top panels are the result in Matsubara space, and the bottom panels are the corresponding results in real frequency. Four temperatures are considered here: $\beta$ = 5 (T $\sim$ 800~K, blue), $\beta$ = 7.5 (T $\sim$ 530~K, orange), $\beta$ = 10 (T $\sim$ 400~K, yellow), and $\beta$ = 15 (T $\sim$ 270~K, purple).

\begin{figure}[tbh]
    \centering
    \includegraphics[width=0.95\columnwidth]{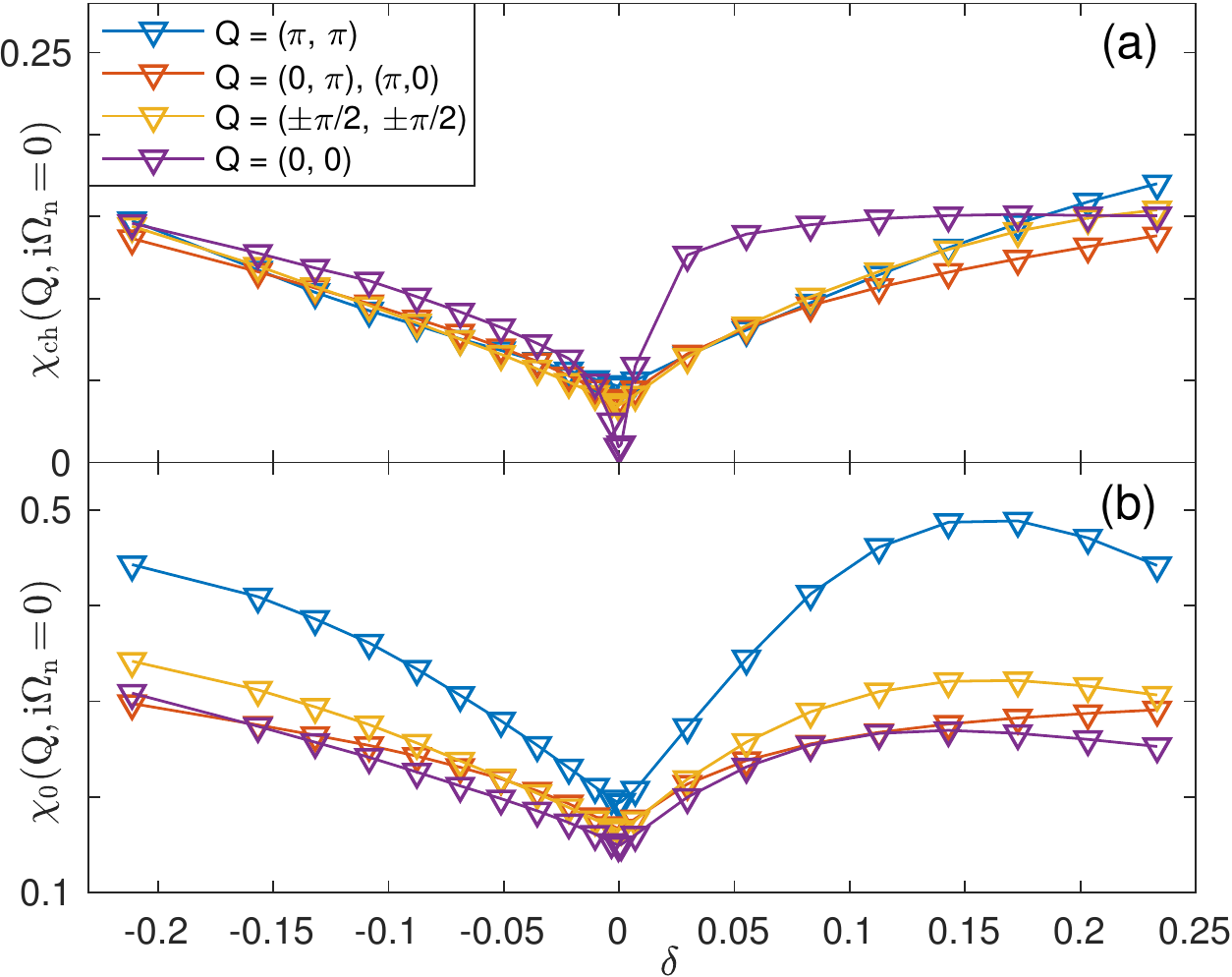}
    \caption{(a) Static charge susceptibility and (b) static charge susceptibility without vertex correction. $\beta = 10$, for different $Q$ at different doping levels.}
    \label{static}
\end{figure}

As temperature at half filling is decreased, low frequency values are reduced, while the high frequency values remain invariant. In real frequency, the peak of the analytically continued data moves from $\Omega \sim 6t$ to $\Omega \sim 7.5t$. The reduction in low frequency values decreases as we increase doping. All other cases (underdoped, optimally doped, and overdoped) do not show much temperature dependence in both Matsubara and real frequency. The inset of [Figure. \ref{doping_difftemp} (b)] shows the temperature dependence of static, uniform [$i\Omega_n=0, Q=(0,0)$] charge susceptibility at doping level $\delta = -0.05$. As shown in Fig. \ref{fig:PDOverview}, at this doping level, we gradually enter the pseudogap regime as temperature is decreased from $\beta = $ 5 to 15. While Ref. \cite{Macridin06} shows that $\chi_m(Q = (0, 0), 0)$ is strongly suppressed at these parameters, $\chi_{ch}$ does not show any signature of the pseudogap. The absence of temperature dependence of $\chi_{ch}$ as we enter the pseudogap regime also means that its doping evolution shown in Fig.~\ref{doping_diffmu} is not caused by the existence of pseudogap.

It is interesting to compare the values from our calculations to those obtained without vertex corrections, where data are obtained by convolving $\chi_0= G * G$. Figure.~\ref{static} shows value for the zero-frequency part, as a function of both electron and hole doping. Remarkably, the vertex corrections strongly suppress the overall charge susceptibility and eliminate a large part of the momentum dependence. In particular, the dominant contribution at $(\pi,\pi)$ is reduced to values similar to the other momenta.

An analysis of the frequency dependence (not shown here) shows the biggest discrepancies at the $\Gamma$ point, where the violation of the charge conservation in the absence of vertex corrections leads to a large frequency-dependent contribution for nonzero momenta.

Our data also shows a pronounced dependence of the charge fluctuations on $t'$. Whereas the hole-doped side just shows a slowly increasing momentum-independent charge susceptibility and an overall suppression of the $(\pi,\pi)$ susceptibility to the level of the other momenta, the electron-doped side (positive $\delta$, note that approximately $\delta \rightarrow -\delta$ for $t'\rightarrow -t'$) shows a large enhancement of the (0, 0) susceptibility as compared to other momenta. This region is close to the onset of a first-order coexistence regime \cite{Macridin06,Gull09_8site,Liebsch09} in this model. Notably this is the same parameter region where strong short-range antiferromagnetic fluctuations are present in the magnetic susceptibility.

Two-particle fluctuations such as the charge fluctuations analyzed in this paper are often interpreted as the underlying cause of changes to single-particle observables. 
From a computational standpoint, both single- and two-particle quantities are computational outcomes of a simulation of Eq.~(\ref{H}). Attributing certain two-particle fluctuations as the ``underlying cause" of a change of single-particle features is therefore difficult. However, as defined in Ref.~\cite{Gunnarsson15}, it is possible to express the single-particle self-energy, and thereby correlation contributions to the change of the spectral functions, in terms of two-particle quantities, via the (exact) equation of motion. For magnetic and charge fluctuations, these equations are

{\small
\begin{align} 
\tilde{\Sigma}_{Q}(K) & = 
 \frac{U}{\beta^2 N_c}\sum_{K'} \, F_{c,m}^{KK'Q} \, G(K')G(K'\! +\!Q)G(K \!+ \!Q)  \\
= & - \, \frac{U}{\beta^2 N_c}\sum_{K'} \, F_{c,d}^{KK'Q} \, G(K')G(K'\!+\!Q)G(K\!+\!Q)
\end{align}}\\
with $F_d$ defined in Eq.~(\ref{eq:chid}), $F_m$ the magnetic analog (for detail mathematical form see Ref.~[\onlinecite{Roheringer12}]), and $K$, $K'$, $Q$ representing pairs of frequency and momentum.

If a single momentum comprises the majority of the self-energy, and contributions from low energies are dominant, then a description in terms of bosonic modes of that type is convenient. This procedure, called ``fluctuation diagnostics", has been successfully applied to non-Fermi liquid \cite{Wu17, LeBlanc19} and real-space correlation functions \cite{Gunnarson18}. 

\begin{figure}[bt]
    \centering
    \includegraphics[width=0.95\columnwidth]{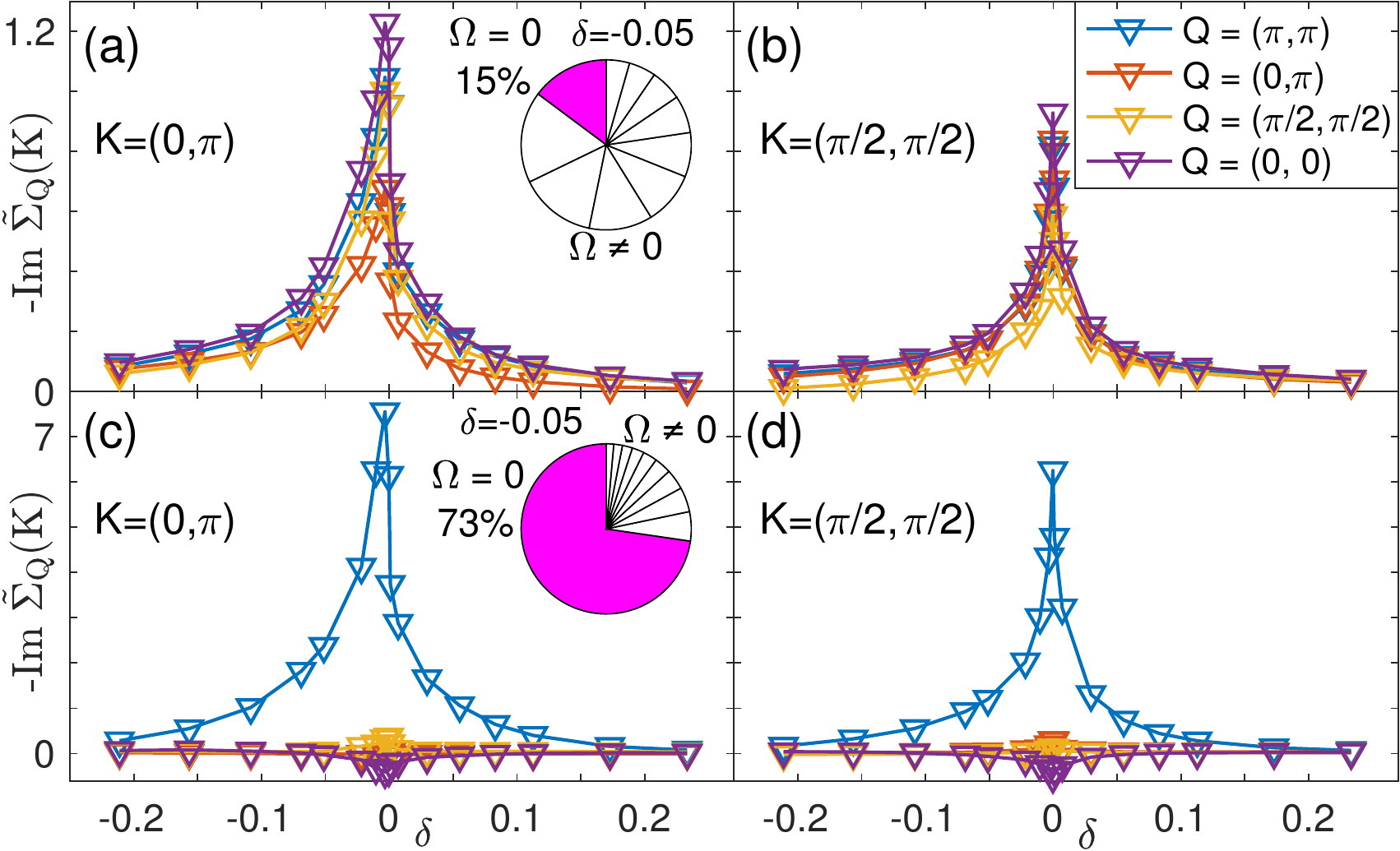}
    \caption{Fluctuation diagnostics \cite{Gunnarsson15} at $\beta = 10$. (a) Charge channel, $K = (0, \pi)$; (b) charge channel, $K = (\pi/2, \pi/2)$. (c) Spin channel, $K = (0, \pi)$; (d) spin channel, $K = (\pi/2, \pi/2)$ as a function of doping. Pi chart: relative magnitude of $|\tilde{\Sigma}_{Q}(K,\pi/\beta)|$ for the first 10 Matsubara frequencies $|\Omega|$ in the charge (a) and spin (c) picture.}
    \label{fluctuation}
\end{figure}

Figure.~\ref{fluctuation} shows the contributions of fluctuations to the single-particle self-energy at the antinode (left panels) and at the node (right panels) expressed in terms of the charge contributions discussed in this paper (top panels) and in terms of magnetic contributions \cite{LeBlanc19_neutron}. As is evident from the lower panels, the pseudogap is well described by short-ranged $Q=(\pi,\pi)$ magnetic fluctuations. A description in terms of charge modes requires similar contributions from all momenta and a much broader frequency range, leading us to conclude that charge fluctuations are not a good way to describe pseudogap physics in the entire parameter range studied here.

In conclusion, we have analyzed the momentum-dependent charge susceptibility in the Hubbard model for a range of dopings and temperatures and for interaction strengths that are thought to be relevant to superconducting cuprates \cite{Gull12_energy,Gull13_super}. Experimental progress in M-EELS \cite{Sean17,Hussian19} and RIXS promises to make this quantity accessible and will provide data that is directly comparable to our results. Our analysis has shown that the dynamical charge susceptibility can be represented by a single peak at a characteristic frequency that exhibits remarkably little momentum dependence, almost no temperature dependence, and a doping dependence that predominantly shifts the peak to lower frequencies as doping is increased. Vertex contributions are essential, as they eliminate the momentum dependence and lower the overall magnitude of the bare susceptibility at $Q=(\pi,\pi)$. As shown by our fluctuation analysis, charge fluctuations are not convenient to understand changes in the single-particle self-energy and spectral function, as terms from all momenta and many frequencies contribute to the self-energy with comparable strength.

While our results are nonperturbative and our solution of the quantum impurity model is numerically exact, the DCA does have important limitations. In particular, due to the limited momentum resolution, DCA is insensitive to stripes with periods larger than our cluster size. Such stripes are seen in experiment \cite{Abbamonte12_stripe, Kohsaka07_stripe, Fujita04_stripe, Parker10, Lawler10_stripe, Hashimoto10_stripe, Daou10_stripe} and theory \cite{White03_stripe,Hager05_stripe,Chang08_strip, Scalapino12_stripe,Zheng16_stripe,Huang17_stripe,Zheng17_stripe} in certain parameter regimes, where they compete with the superconducting state, and have been proposed as the cause or a consequence of the pseudogap \cite{Lawler10_stripe, Hashimoto10_stripe, Daou10_stripe, Parker10}.

\begin{acknowledgments}
This work was supported by the National Science Foundation under Grant No. DMR-1606348. This work used resources of the Extreme Science and Engineering Discovering Environment (XSEDE) under Grant No. TG-DMR130036. The Flatiron Institute is a division of the Simons Foundation.
\end{acknowledgments}

\bibliographystyle{apsrev4-1}
\bibliography{refs}

\begin{thebibliography}{84}%
\makeatletter
\providecommand \@ifxundefined [1]{%
 \@ifx{#1\undefined}
}%
\providecommand \@ifnum [1]{%
 \ifnum #1\expandafter \@firstoftwo
 \else \expandafter \@secondoftwo
 \fi
}%
\providecommand \@ifx [1]{%
 \ifx #1\expandafter \@firstoftwo
 \else \expandafter \@secondoftwo
 \fi
}%
\providecommand \natexlab [1]{#1}%
\providecommand \enquote  [1]{``#1''}%
\providecommand \bibnamefont  [1]{#1}%
\providecommand \bibfnamefont [1]{#1}%
\providecommand \citenamefont [1]{#1}%
\providecommand \href@noop [0]{\@secondoftwo}%
\providecommand \href [0]{\begingroup \@sanitize@url \@href}%
\providecommand \@href[1]{\@@startlink{#1}\@@href}%
\providecommand \@@href[1]{\endgroup#1\@@endlink}%
\providecommand \@sanitize@url [0]{\catcode `\\12\catcode `\$12\catcode
  `\&12\catcode `\#12\catcode `\^12\catcode `\_12\catcode `\%12\relax}%
\providecommand \@@startlink[1]{}%
\providecommand \@@endlink[0]{}%
\providecommand \url  [0]{\begingroup\@sanitize@url \@url }%
\providecommand \@url [1]{\endgroup\@href {#1}{\urlprefix }}%
\providecommand \urlprefix  [0]{URL }%
\providecommand \Eprint [0]{\href }%
\providecommand \doibase [0]{http://dx.doi.org/}%
\providecommand \selectlanguage [0]{\@gobble}%
\providecommand \bibinfo  [0]{\@secondoftwo}%
\providecommand \bibfield  [0]{\@secondoftwo}%
\providecommand \translation [1]{[#1]}%
\providecommand \BibitemOpen [0]{}%
\providecommand \bibitemStop [0]{}%
\providecommand \bibitemNoStop [0]{.\EOS\space}%
\providecommand \EOS [0]{\spacefactor3000\relax}%
\providecommand \BibitemShut  [1]{\csname bibitem#1\endcsname}%
\let\auto@bib@innerbib\@empty
\bibitem [{\citenamefont {Ishii}\ \emph {et~al.}(2017)\citenamefont {Ishii},
  \citenamefont {Tohyama}, \citenamefont {Asano}, \citenamefont {Sato},
  \citenamefont {Fujita}, \citenamefont {Wakimoto}, \citenamefont {Tustsui},
  \citenamefont {Sota}, \citenamefont {Miyawaki}, \citenamefont {Niwa},
  \citenamefont {Harada}, \citenamefont {Pelliciari}, \citenamefont {Huang},
  \citenamefont {Schmitt}, \citenamefont {Yamamoto},\ and\ \citenamefont
  {Mizuki}}]{Ishii17}%
  \BibitemOpen
  \bibfield  {author} {\bibinfo {author} {\bibfnamefont {K.}~\bibnamefont
  {Ishii}}, \bibinfo {author} {\bibfnamefont {T.}~\bibnamefont {Tohyama}},
  \bibinfo {author} {\bibfnamefont {S.}~\bibnamefont {Asano}}, \bibinfo
  {author} {\bibfnamefont {K.}~\bibnamefont {Sato}}, \bibinfo {author}
  {\bibfnamefont {M.}~\bibnamefont {Fujita}}, \bibinfo {author} {\bibfnamefont
  {S.}~\bibnamefont {Wakimoto}}, \bibinfo {author} {\bibfnamefont
  {K.}~\bibnamefont {Tustsui}}, \bibinfo {author} {\bibfnamefont
  {S.}~\bibnamefont {Sota}}, \bibinfo {author} {\bibfnamefont {J.}~\bibnamefont
  {Miyawaki}}, \bibinfo {author} {\bibfnamefont {H.}~\bibnamefont {Niwa}},
  \bibinfo {author} {\bibfnamefont {Y.}~\bibnamefont {Harada}}, \bibinfo
  {author} {\bibfnamefont {J.}~\bibnamefont {Pelliciari}}, \bibinfo {author}
  {\bibfnamefont {Y.}~\bibnamefont {Huang}}, \bibinfo {author} {\bibfnamefont
  {T.}~\bibnamefont {Schmitt}}, \bibinfo {author} {\bibfnamefont
  {Y.}~\bibnamefont {Yamamoto}}, \ and\ \bibinfo {author} {\bibfnamefont
  {J.}~\bibnamefont {Mizuki}},\ }\href {\doibase 10.1103/PhysRevB.96.115148}
  {\bibfield  {journal} {\bibinfo  {journal} {Phys. Rev. B}\ }\textbf {\bibinfo
  {volume} {96}},\ \bibinfo {pages} {115148} (\bibinfo {year}
  {2017})}\BibitemShut {NoStop}%
\bibitem [{\citenamefont {Suzuki}\ \emph {et~al.}(2018)\citenamefont {Suzuki},
  \citenamefont {Minola}, \citenamefont {Lu}, \citenamefont {Peng},
  \citenamefont {Fumagalli}, \citenamefont {Lefran{\c{c}}ois}, \citenamefont
  {Loew}, \citenamefont {Porras}, \citenamefont {Kummer}, \citenamefont
  {Betto}, \citenamefont {Ishida}, \citenamefont {Eisaki}, \citenamefont {Hu},
  \citenamefont {Zhou}, \citenamefont {Haverkort}, \citenamefont {Brookes},
  \citenamefont {Braicovich}, \citenamefont {Ghiringhelli}, \citenamefont
  {Le~Tacon},\ and\ \citenamefont {Keimer}}]{Suzuki18}%
  \BibitemOpen
  \bibfield  {author} {\bibinfo {author} {\bibfnamefont {H.}~\bibnamefont
  {Suzuki}}, \bibinfo {author} {\bibfnamefont {M.}~\bibnamefont {Minola}},
  \bibinfo {author} {\bibfnamefont {Y.}~\bibnamefont {Lu}}, \bibinfo {author}
  {\bibfnamefont {Y.}~\bibnamefont {Peng}}, \bibinfo {author} {\bibfnamefont
  {R.}~\bibnamefont {Fumagalli}}, \bibinfo {author} {\bibfnamefont
  {E.}~\bibnamefont {Lefran{\c{c}}ois}}, \bibinfo {author} {\bibfnamefont
  {T.}~\bibnamefont {Loew}}, \bibinfo {author} {\bibfnamefont {J.}~\bibnamefont
  {Porras}}, \bibinfo {author} {\bibfnamefont {K.}~\bibnamefont {Kummer}},
  \bibinfo {author} {\bibfnamefont {D.}~\bibnamefont {Betto}}, \bibinfo
  {author} {\bibfnamefont {S.}~\bibnamefont {Ishida}}, \bibinfo {author}
  {\bibfnamefont {H.}~\bibnamefont {Eisaki}}, \bibinfo {author} {\bibfnamefont
  {C.}~\bibnamefont {Hu}}, \bibinfo {author} {\bibfnamefont {X.}~\bibnamefont
  {Zhou}}, \bibinfo {author} {\bibfnamefont {M.~W.}\ \bibnamefont {Haverkort}},
  \bibinfo {author} {\bibfnamefont {N.~B.}\ \bibnamefont {Brookes}}, \bibinfo
  {author} {\bibfnamefont {L.}~\bibnamefont {Braicovich}}, \bibinfo {author}
  {\bibfnamefont {G.}~\bibnamefont {Ghiringhelli}}, \bibinfo {author}
  {\bibfnamefont {M.}~\bibnamefont {Le~Tacon}}, \ and\ \bibinfo {author}
  {\bibfnamefont {B.}~\bibnamefont {Keimer}},\ }\href {\doibase
  10.1038/s41535-018-0139-7} {\bibfield  {journal} {\bibinfo  {journal} {npj
  Quantum Materials}\ }\textbf {\bibinfo {volume} {3}},\ \bibinfo {pages} {65}
  (\bibinfo {year} {2018})}\BibitemShut {NoStop}%
\bibitem [{\citenamefont {da~Silva~Neto}\ \emph {et~al.}(2018)\citenamefont
  {da~Silva~Neto}, \citenamefont {Minola}, \citenamefont {Yu}, \citenamefont
  {Tabis}, \citenamefont {Bluschke}, \citenamefont {Unruh}, \citenamefont
  {Suzuki}, \citenamefont {Li}, \citenamefont {Yu}, \citenamefont {Betto},
  \citenamefont {Kummer}, \citenamefont {Yakhou}, \citenamefont {Brookes},
  \citenamefont {Le~Tacon}, \citenamefont {Greven}, \citenamefont {Keimer},\
  and\ \citenamefont {Damascelli}}]{Neto18}%
  \BibitemOpen
  \bibfield  {author} {\bibinfo {author} {\bibfnamefont {E.~H.}\ \bibnamefont
  {da~Silva~Neto}}, \bibinfo {author} {\bibfnamefont {M.}~\bibnamefont
  {Minola}}, \bibinfo {author} {\bibfnamefont {B.}~\bibnamefont {Yu}}, \bibinfo
  {author} {\bibfnamefont {W.}~\bibnamefont {Tabis}}, \bibinfo {author}
  {\bibfnamefont {M.}~\bibnamefont {Bluschke}}, \bibinfo {author}
  {\bibfnamefont {D.}~\bibnamefont {Unruh}}, \bibinfo {author} {\bibfnamefont
  {H.}~\bibnamefont {Suzuki}}, \bibinfo {author} {\bibfnamefont
  {Y.}~\bibnamefont {Li}}, \bibinfo {author} {\bibfnamefont {G.}~\bibnamefont
  {Yu}}, \bibinfo {author} {\bibfnamefont {D.}~\bibnamefont {Betto}}, \bibinfo
  {author} {\bibfnamefont {K.}~\bibnamefont {Kummer}}, \bibinfo {author}
  {\bibfnamefont {F.}~\bibnamefont {Yakhou}}, \bibinfo {author} {\bibfnamefont
  {N.~B.}\ \bibnamefont {Brookes}}, \bibinfo {author} {\bibfnamefont
  {M.}~\bibnamefont {Le~Tacon}}, \bibinfo {author} {\bibfnamefont
  {M.}~\bibnamefont {Greven}}, \bibinfo {author} {\bibfnamefont
  {B.}~\bibnamefont {Keimer}}, \ and\ \bibinfo {author} {\bibfnamefont
  {A.}~\bibnamefont {Damascelli}},\ }\href {\doibase
  10.1103/PhysRevB.98.161114} {\bibfield  {journal} {\bibinfo  {journal} {Phys.
  Rev. B}\ }\textbf {\bibinfo {volume} {98}},\ \bibinfo {pages} {161114(R)}
  (\bibinfo {year} {2018})}\BibitemShut {NoStop}%
\bibitem [{\citenamefont {Hepting}\ \emph {et~al.}(2018)\citenamefont
  {Hepting}, \citenamefont {Chaix}, \citenamefont {Huang}, \citenamefont
  {Fumagalli}, \citenamefont {Peng}, \citenamefont {Moritz}, \citenamefont
  {Kummer}, \citenamefont {Brookes}, \citenamefont {Lee}, \citenamefont
  {Hashimoto}, \citenamefont {Sarkar}, \citenamefont {He}, \citenamefont
  {Rotundu}, \citenamefont {Lee}, \citenamefont {Greene}, \citenamefont
  {Braicovich}, \citenamefont {Ghiringhelli}, \citenamefont {Shen},
  \citenamefont {Devereaux},\ and\ \citenamefont {Lee}}]{Hepting18}%
  \BibitemOpen
  \bibfield  {author} {\bibinfo {author} {\bibfnamefont {M.}~\bibnamefont
  {Hepting}}, \bibinfo {author} {\bibfnamefont {L.}~\bibnamefont {Chaix}},
  \bibinfo {author} {\bibfnamefont {E.~W.}\ \bibnamefont {Huang}}, \bibinfo
  {author} {\bibfnamefont {R.}~\bibnamefont {Fumagalli}}, \bibinfo {author}
  {\bibfnamefont {Y.~Y.}\ \bibnamefont {Peng}}, \bibinfo {author}
  {\bibfnamefont {B.}~\bibnamefont {Moritz}}, \bibinfo {author} {\bibfnamefont
  {K.}~\bibnamefont {Kummer}}, \bibinfo {author} {\bibfnamefont {N.~B.}\
  \bibnamefont {Brookes}}, \bibinfo {author} {\bibfnamefont {W.~C.}\
  \bibnamefont {Lee}}, \bibinfo {author} {\bibfnamefont {M.}~\bibnamefont
  {Hashimoto}}, \bibinfo {author} {\bibfnamefont {T.}~\bibnamefont {Sarkar}},
  \bibinfo {author} {\bibfnamefont {J.-F.}\ \bibnamefont {He}}, \bibinfo
  {author} {\bibfnamefont {C.~R.}\ \bibnamefont {Rotundu}}, \bibinfo {author}
  {\bibfnamefont {Y.~S.}\ \bibnamefont {Lee}}, \bibinfo {author} {\bibfnamefont
  {R.~L.}\ \bibnamefont {Greene}}, \bibinfo {author} {\bibfnamefont
  {L.}~\bibnamefont {Braicovich}}, \bibinfo {author} {\bibfnamefont
  {G.}~\bibnamefont {Ghiringhelli}}, \bibinfo {author} {\bibfnamefont {Z.~X.}\
  \bibnamefont {Shen}}, \bibinfo {author} {\bibfnamefont {T.~P.}\ \bibnamefont
  {Devereaux}}, \ and\ \bibinfo {author} {\bibfnamefont {W.~S.}\ \bibnamefont
  {Lee}},\ }\href {\doibase 10.1038/s41586-018-0648-3} {\bibfield  {journal}
  {\bibinfo  {journal} {Nature}\ }\textbf {\bibinfo {volume} {563}},\ \bibinfo
  {pages} {374} (\bibinfo {year} {2018})}\BibitemShut {NoStop}%
\bibitem [{\citenamefont {Miao}\ \emph {et~al.}(2019)\citenamefont {Miao},
  \citenamefont {Fumagalli}, \citenamefont {Rossi}, \citenamefont {Lorenzana},
  \citenamefont {Seibold}, \citenamefont {Yakhou-Harris}, \citenamefont
  {Kummer}, \citenamefont {Brookes}, \citenamefont {Gu}, \citenamefont
  {Braicovich}, \citenamefont {Ghiringhelli},\ and\ \citenamefont
  {Dean}}]{Miao19}%
  \BibitemOpen
  \bibfield  {author} {\bibinfo {author} {\bibfnamefont {H.}~\bibnamefont
  {Miao}}, \bibinfo {author} {\bibfnamefont {R.}~\bibnamefont {Fumagalli}},
  \bibinfo {author} {\bibfnamefont {M.}~\bibnamefont {Rossi}}, \bibinfo
  {author} {\bibfnamefont {J.}~\bibnamefont {Lorenzana}}, \bibinfo {author}
  {\bibfnamefont {G.}~\bibnamefont {Seibold}}, \bibinfo {author} {\bibfnamefont
  {F.}~\bibnamefont {Yakhou-Harris}}, \bibinfo {author} {\bibfnamefont
  {K.}~\bibnamefont {Kummer}}, \bibinfo {author} {\bibfnamefont {N.~B.}\
  \bibnamefont {Brookes}}, \bibinfo {author} {\bibfnamefont {G.~D.}\
  \bibnamefont {Gu}}, \bibinfo {author} {\bibfnamefont {L.}~\bibnamefont
  {Braicovich}}, \bibinfo {author} {\bibfnamefont {G.}~\bibnamefont
  {Ghiringhelli}}, \ and\ \bibinfo {author} {\bibfnamefont {M.~P.~M.}\
  \bibnamefont {Dean}},\ }\href {\doibase 10.1103/PhysRevX.9.031042} {\bibfield
   {journal} {\bibinfo  {journal} {Phys. Rev. X}\ }\textbf {\bibinfo {volume}
  {9}},\ \bibinfo {pages} {031042} (\bibinfo {year} {2019})}\BibitemShut
  {NoStop}%
\bibitem [{\citenamefont {Ishii}\ \emph {et~al.}(2019)\citenamefont {Ishii},
  \citenamefont {Kurooka}, \citenamefont {Shimizu}, \citenamefont {Fujita},
  \citenamefont {Yamada},\ and\ \citenamefont {Mizuki}}]{Ishii19}%
  \BibitemOpen
  \bibfield  {author} {\bibinfo {author} {\bibfnamefont {K.}~\bibnamefont
  {Ishii}}, \bibinfo {author} {\bibfnamefont {M.}~\bibnamefont {Kurooka}},
  \bibinfo {author} {\bibfnamefont {Y.}~\bibnamefont {Shimizu}}, \bibinfo
  {author} {\bibfnamefont {M.}~\bibnamefont {Fujita}}, \bibinfo {author}
  {\bibfnamefont {K.}~\bibnamefont {Yamada}}, \ and\ \bibinfo {author}
  {\bibfnamefont {J.}~\bibnamefont {Mizuki}},\ }\href {\doibase
  10.7566/JPSJ.88.075001} {\bibfield  {journal} {\bibinfo  {journal} {Journal
  of the Physical Society of Japan}\ }\textbf {\bibinfo {volume} {88}},\
  \bibinfo {pages} {075001} (\bibinfo {year} {2019})}\BibitemShut {NoStop}%
\bibitem [{\citenamefont {Vig}\ \emph {et~al.}(2017)\citenamefont {Vig},
  \citenamefont {Kogar}, \citenamefont {Mitrano}, \citenamefont {Husain},
  \citenamefont {Mishra}, \citenamefont {Rak}, \citenamefont {Venema},
  \citenamefont {Johnson}, \citenamefont {Gu}, \citenamefont {Fradkin},
  \citenamefont {Norman},\ and\ \citenamefont {Abbamonte}}]{Sean17}%
  \BibitemOpen
  \bibfield  {author} {\bibinfo {author} {\bibfnamefont {S.}~\bibnamefont
  {Vig}}, \bibinfo {author} {\bibfnamefont {A.}~\bibnamefont {Kogar}}, \bibinfo
  {author} {\bibfnamefont {M.}~\bibnamefont {Mitrano}}, \bibinfo {author}
  {\bibfnamefont {A.~A.}\ \bibnamefont {Husain}}, \bibinfo {author}
  {\bibfnamefont {V.}~\bibnamefont {Mishra}}, \bibinfo {author} {\bibfnamefont
  {M.~S.}\ \bibnamefont {Rak}}, \bibinfo {author} {\bibfnamefont
  {L.}~\bibnamefont {Venema}}, \bibinfo {author} {\bibfnamefont {P.~D.}\
  \bibnamefont {Johnson}}, \bibinfo {author} {\bibfnamefont {G.~D.}\
  \bibnamefont {Gu}}, \bibinfo {author} {\bibfnamefont {E.}~\bibnamefont
  {Fradkin}}, \bibinfo {author} {\bibfnamefont {M.~R.}\ \bibnamefont {Norman}},
  \ and\ \bibinfo {author} {\bibfnamefont {P.}~\bibnamefont {Abbamonte}},\
  }\href {\doibase 10.21468/SciPostPhys.3.4.026} {\bibfield  {journal}
  {\bibinfo  {journal} {SciPost Phys.}\ }\textbf {\bibinfo {volume} {3}},\
  \bibinfo {pages} {026} (\bibinfo {year} {2017})}\BibitemShut {NoStop}%
\bibitem [{\citenamefont {{Husain}}\ \emph {et~al.}(2019)\citenamefont
  {{Husain}}, \citenamefont {{Mitrano}}, \citenamefont {{Rak}}, \citenamefont
  {{Rubeck}}, \citenamefont {{Uchoa}}, \citenamefont {{Schneeloch}},
  \citenamefont {{Zhong}}, \citenamefont {{Gu}},\ and\ \citenamefont
  {{Abbamonte}}}]{Hussian19}%
  \BibitemOpen
  \bibfield  {author} {\bibinfo {author} {\bibfnamefont {A.~A.}\ \bibnamefont
  {{Husain}}}, \bibinfo {author} {\bibfnamefont {M.}~\bibnamefont {{Mitrano}}},
  \bibinfo {author} {\bibfnamefont {M.~S.}\ \bibnamefont {{Rak}}}, \bibinfo
  {author} {\bibfnamefont {S.~I.}\ \bibnamefont {{Rubeck}}}, \bibinfo {author}
  {\bibfnamefont {B.}~\bibnamefont {{Uchoa}}}, \bibinfo {author} {\bibfnamefont
  {J.}~\bibnamefont {{Schneeloch}}}, \bibinfo {author} {\bibfnamefont
  {R.}~\bibnamefont {{Zhong}}}, \bibinfo {author} {\bibfnamefont {G.~D.}\
  \bibnamefont {{Gu}}}, \ and\ \bibinfo {author} {\bibfnamefont
  {P.}~\bibnamefont {{Abbamonte}}},\ }\href@noop {} {\enquote {\bibinfo {title}
  {{Crossover of Charge Fluctuations across the Strange Metal Phase
  Diagram}},}\ } (\bibinfo {year} {2019}),\ \Eprint
  {http://arxiv.org/abs/1903.04038} {arXiv:1903.04038} \BibitemShut {NoStop}%
\bibitem [{\citenamefont {Anderson}(1987)}]{Anderson87}%
  \BibitemOpen
  \bibfield  {author} {\bibinfo {author} {\bibfnamefont {P.~W.}\ \bibnamefont
  {Anderson}},\ }\href {\doibase 10.1126/science.235.4793.1196} {\bibfield
  {journal} {\bibinfo  {journal} {Science}\ }\textbf {\bibinfo {volume}
  {235}},\ \bibinfo {pages} {1196} (\bibinfo {year} {1987})}\BibitemShut
  {NoStop}%
\bibitem [{\citenamefont {Scalapino}(2007)}]{Scalapino07}%
  \BibitemOpen
  \bibfield  {author} {\bibinfo {author} {\bibfnamefont {D.}~\bibnamefont
  {Scalapino}},\ }in\ \href@noop {} {\emph {\bibinfo {booktitle} {Handbook of
  High-Temperature Superconductivity}}},\ \bibinfo {editor} {edited by\
  \bibinfo {editor} {\bibfnamefont {J.}~\bibnamefont {Schrieffer}}\ and\
  \bibinfo {editor} {\bibfnamefont {J.}~\bibnamefont {Brooks}}}\ (\bibinfo
  {publisher} {Springer New York},\ \bibinfo {year} {2007})\ pp.\ \bibinfo
  {pages} {495--526}\BibitemShut {NoStop}%
\bibitem [{\citenamefont {Huscroft}\ \emph {et~al.}(2001)\citenamefont
  {Huscroft}, \citenamefont {Jarrell}, \citenamefont {Maier}, \citenamefont
  {Moukouri},\ and\ \citenamefont {Tahvildarzadeh}}]{Huscroft01}%
  \BibitemOpen
  \bibfield  {author} {\bibinfo {author} {\bibfnamefont {C.}~\bibnamefont
  {Huscroft}}, \bibinfo {author} {\bibfnamefont {M.}~\bibnamefont {Jarrell}},
  \bibinfo {author} {\bibfnamefont {T.}~\bibnamefont {Maier}}, \bibinfo
  {author} {\bibfnamefont {S.}~\bibnamefont {Moukouri}}, \ and\ \bibinfo
  {author} {\bibfnamefont {A.~N.}\ \bibnamefont {Tahvildarzadeh}},\ }\href
  {\doibase 10.1103/PhysRevLett.86.139} {\bibfield  {journal} {\bibinfo
  {journal} {Phys. Rev. Lett.}\ }\textbf {\bibinfo {volume} {86}},\ \bibinfo
  {pages} {139} (\bibinfo {year} {2001})}\BibitemShut {NoStop}%
\bibitem [{\citenamefont {Civelli}\ \emph {et~al.}(2005)\citenamefont
  {Civelli}, \citenamefont {Capone}, \citenamefont {Kancharla}, \citenamefont
  {Parcollet},\ and\ \citenamefont {Kotliar}}]{Civelli05}%
  \BibitemOpen
  \bibfield  {author} {\bibinfo {author} {\bibfnamefont {M.}~\bibnamefont
  {Civelli}}, \bibinfo {author} {\bibfnamefont {M.}~\bibnamefont {Capone}},
  \bibinfo {author} {\bibfnamefont {S.~S.}\ \bibnamefont {Kancharla}}, \bibinfo
  {author} {\bibfnamefont {O.}~\bibnamefont {Parcollet}}, \ and\ \bibinfo
  {author} {\bibfnamefont {G.}~\bibnamefont {Kotliar}},\ }\href {\doibase
  10.1103/PhysRevLett.95.106402} {\bibfield  {journal} {\bibinfo  {journal}
  {Phys. Rev. Lett.}\ }\textbf {\bibinfo {volume} {95}},\ \bibinfo {eid}
  {106402} (\bibinfo {year} {2005})}\BibitemShut {NoStop}%
\bibitem [{\citenamefont {Stanescu}\ and\ \citenamefont
  {Kotliar}(2006)}]{Stanescu06}%
  \BibitemOpen
  \bibfield  {author} {\bibinfo {author} {\bibfnamefont {T.~D.}\ \bibnamefont
  {Stanescu}}\ and\ \bibinfo {author} {\bibfnamefont {G.}~\bibnamefont
  {Kotliar}},\ }\href {\doibase 10.1103/PhysRevB.74.125110} {\bibfield
  {journal} {\bibinfo  {journal} {Phys. Rev. B}\ }\textbf {\bibinfo {volume}
  {74}},\ \bibinfo {pages} {125110} (\bibinfo {year} {2006})}\BibitemShut
  {NoStop}%
\bibitem [{\citenamefont {Park}\ \emph {et~al.}(2008)\citenamefont {Park},
  \citenamefont {Haule},\ and\ \citenamefont {Kotliar}}]{Parker08}%
  \BibitemOpen
  \bibfield  {author} {\bibinfo {author} {\bibfnamefont {H.}~\bibnamefont
  {Park}}, \bibinfo {author} {\bibfnamefont {K.}~\bibnamefont {Haule}}, \ and\
  \bibinfo {author} {\bibfnamefont {G.}~\bibnamefont {Kotliar}},\ }\href
  {\doibase 10.1103/PhysRevLett.101.186403} {\bibfield  {journal} {\bibinfo
  {journal} {Phys. Rev. Lett.}\ }\textbf {\bibinfo {volume} {101}},\ \bibinfo
  {pages} {186403} (\bibinfo {year} {2008})}\BibitemShut {NoStop}%
\bibitem [{\citenamefont {Liebsch}\ and\ \citenamefont
  {Tong}(2009)}]{Liebsch09}%
  \BibitemOpen
  \bibfield  {author} {\bibinfo {author} {\bibfnamefont {A.}~\bibnamefont
  {Liebsch}}\ and\ \bibinfo {author} {\bibfnamefont {N.-H.}\ \bibnamefont
  {Tong}},\ }\href {\doibase 10.1103/PhysRevB.80.165126} {\bibfield  {journal}
  {\bibinfo  {journal} {Phys. Rev. B}\ }\textbf {\bibinfo {volume} {80}},\
  \bibinfo {pages} {165126} (\bibinfo {year} {2009})}\BibitemShut {NoStop}%
\bibitem [{\citenamefont {Ferrero}\ \emph
  {et~al.}(2009{\natexlab{a}})\citenamefont {Ferrero}, \citenamefont
  {Cornaglia}, \citenamefont {De~Leo}, \citenamefont {Parcollet}, \citenamefont
  {Kotliar},\ and\ \citenamefont {Georges}}]{Ferrero09}%
  \BibitemOpen
  \bibfield  {author} {\bibinfo {author} {\bibfnamefont {M.}~\bibnamefont
  {Ferrero}}, \bibinfo {author} {\bibfnamefont {P.~S.}\ \bibnamefont
  {Cornaglia}}, \bibinfo {author} {\bibfnamefont {L.}~\bibnamefont {De~Leo}},
  \bibinfo {author} {\bibfnamefont {O.}~\bibnamefont {Parcollet}}, \bibinfo
  {author} {\bibfnamefont {G.}~\bibnamefont {Kotliar}}, \ and\ \bibinfo
  {author} {\bibfnamefont {A.}~\bibnamefont {Georges}},\ }\href {\doibase
  10.1103/PhysRevB.80.064501} {\bibfield  {journal} {\bibinfo  {journal} {Phys.
  Rev. B}\ }\textbf {\bibinfo {volume} {80}},\ \bibinfo {pages} {064501}
  (\bibinfo {year} {2009}{\natexlab{a}})}\BibitemShut {NoStop}%
\bibitem [{\citenamefont {Ferrero}\ \emph
  {et~al.}(2009{\natexlab{b}})\citenamefont {Ferrero}, \citenamefont
  {Cornaglia}, \citenamefont {Leo}, \citenamefont {Parcollet}, \citenamefont
  {Kotliar},\ and\ \citenamefont {Georges}}]{Ferrero09B}%
  \BibitemOpen
  \bibfield  {author} {\bibinfo {author} {\bibfnamefont {M.}~\bibnamefont
  {Ferrero}}, \bibinfo {author} {\bibfnamefont {P.~S.}\ \bibnamefont
  {Cornaglia}}, \bibinfo {author} {\bibfnamefont {L.~D.}\ \bibnamefont {Leo}},
  \bibinfo {author} {\bibfnamefont {O.}~\bibnamefont {Parcollet}}, \bibinfo
  {author} {\bibfnamefont {G.}~\bibnamefont {Kotliar}}, \ and\ \bibinfo
  {author} {\bibfnamefont {A.}~\bibnamefont {Georges}},\ }\href {\doibase
  10.1209/0295-5075/85/57009} {\bibfield  {journal} {\bibinfo  {journal} {EPL
  (Europhysics Letters)}\ }\textbf {\bibinfo {volume} {85}},\ \bibinfo {pages}
  {57009} (\bibinfo {year} {2009}{\natexlab{b}})}\BibitemShut {NoStop}%
\bibitem [{\citenamefont {Sakai}\ \emph {et~al.}(2009)\citenamefont {Sakai},
  \citenamefont {Motome},\ and\ \citenamefont {Imada}}]{Sakai09}%
  \BibitemOpen
  \bibfield  {author} {\bibinfo {author} {\bibfnamefont {S.}~\bibnamefont
  {Sakai}}, \bibinfo {author} {\bibfnamefont {Y.}~\bibnamefont {Motome}}, \
  and\ \bibinfo {author} {\bibfnamefont {M.}~\bibnamefont {Imada}},\ }\href
  {\doibase 10.1103/PhysRevLett.102.056404} {\bibfield  {journal} {\bibinfo
  {journal} {Phys. Rev. Lett.}\ }\textbf {\bibinfo {volume} {102}},\ \bibinfo
  {pages} {056404} (\bibinfo {year} {2009})}\BibitemShut {NoStop}%
\bibitem [{\citenamefont {Sakai}\ \emph {et~al.}(2010)\citenamefont {Sakai},
  \citenamefont {Motome},\ and\ \citenamefont {Imada}}]{Sakai10}%
  \BibitemOpen
  \bibfield  {author} {\bibinfo {author} {\bibfnamefont {S.}~\bibnamefont
  {Sakai}}, \bibinfo {author} {\bibfnamefont {Y.}~\bibnamefont {Motome}}, \
  and\ \bibinfo {author} {\bibfnamefont {M.}~\bibnamefont {Imada}},\ }\href
  {\doibase 10.1103/PhysRevB.82.134505} {\bibfield  {journal} {\bibinfo
  {journal} {Phys. Rev. B}\ }\textbf {\bibinfo {volume} {82}},\ \bibinfo
  {pages} {134505} (\bibinfo {year} {2010})}\BibitemShut {NoStop}%
\bibitem [{\citenamefont {Lin}\ \emph {et~al.}(2010)\citenamefont {Lin},
  \citenamefont {Gull},\ and\ \citenamefont {Millis}}]{Lin10}%
  \BibitemOpen
  \bibfield  {author} {\bibinfo {author} {\bibfnamefont {N.}~\bibnamefont
  {Lin}}, \bibinfo {author} {\bibfnamefont {E.}~\bibnamefont {Gull}}, \ and\
  \bibinfo {author} {\bibfnamefont {A.~J.}\ \bibnamefont {Millis}},\ }\href
  {\doibase 10.1103/PhysRevB.82.045104} {\bibfield  {journal} {\bibinfo
  {journal} {Phys. Rev. B}\ }\textbf {\bibinfo {volume} {82}},\ \bibinfo
  {pages} {045104} (\bibinfo {year} {2010})}\BibitemShut {NoStop}%
\bibitem [{\citenamefont {Ioffe}\ and\ \citenamefont {Millis}(2000)}]{Ioffe00}%
  \BibitemOpen
  \bibfield  {author} {\bibinfo {author} {\bibfnamefont {L.~B.}\ \bibnamefont
  {Ioffe}}\ and\ \bibinfo {author} {\bibfnamefont {A.~J.}\ \bibnamefont
  {Millis}},\ }\href {\doibase 10.1103/PhysRevB.61.9077} {\bibfield  {journal}
  {\bibinfo  {journal} {Phys. Rev. B}\ }\textbf {\bibinfo {volume} {61}},\
  \bibinfo {pages} {9077} (\bibinfo {year} {2000})}\BibitemShut {NoStop}%
\bibitem [{\citenamefont {Toschi}\ \emph {et~al.}(2005)\citenamefont {Toschi},
  \citenamefont {Capone},\ and\ \citenamefont {Castellani}}]{Toschi05}%
  \BibitemOpen
  \bibfield  {author} {\bibinfo {author} {\bibfnamefont {A.}~\bibnamefont
  {Toschi}}, \bibinfo {author} {\bibfnamefont {M.}~\bibnamefont {Capone}}, \
  and\ \bibinfo {author} {\bibfnamefont {C.}~\bibnamefont {Castellani}},\
  }\href {\doibase 10.1103/PhysRevB.72.235118} {\bibfield  {journal} {\bibinfo
  {journal} {Phys. Rev. B}\ }\textbf {\bibinfo {volume} {72}},\ \bibinfo
  {pages} {235118} (\bibinfo {year} {2005})}\BibitemShut {NoStop}%
\bibitem [{\citenamefont {Millis}\ \emph {et~al.}(2005)\citenamefont {Millis},
  \citenamefont {Zimmers}, \citenamefont {Lobo}, \citenamefont {Bontemps},\
  and\ \citenamefont {Homes}}]{Millis05}%
  \BibitemOpen
  \bibfield  {author} {\bibinfo {author} {\bibfnamefont {A.~J.}\ \bibnamefont
  {Millis}}, \bibinfo {author} {\bibfnamefont {A.}~\bibnamefont {Zimmers}},
  \bibinfo {author} {\bibfnamefont {R.~P. S.~M.}\ \bibnamefont {Lobo}},
  \bibinfo {author} {\bibfnamefont {N.}~\bibnamefont {Bontemps}}, \ and\
  \bibinfo {author} {\bibfnamefont {C.~C.}\ \bibnamefont {Homes}},\ }\href
  {\doibase 10.1103/PhysRevB.72.224517} {\bibfield  {journal} {\bibinfo
  {journal} {Phys. Rev. B}\ }\textbf {\bibinfo {volume} {72}},\ \bibinfo
  {pages} {224517} (\bibinfo {year} {2005})}\BibitemShut {NoStop}%
\bibitem [{\citenamefont {Comanac}\ \emph {et~al.}(2008)\citenamefont
  {Comanac}, \citenamefont {de' Medici}, \citenamefont {Capone},\ and\
  \citenamefont {Millis}}]{Comanac08}%
  \BibitemOpen
  \bibfield  {author} {\bibinfo {author} {\bibfnamefont {A.}~\bibnamefont
  {Comanac}}, \bibinfo {author} {\bibfnamefont {L.}~\bibnamefont {de' Medici}},
  \bibinfo {author} {\bibfnamefont {M.}~\bibnamefont {Capone}}, \ and\ \bibinfo
  {author} {\bibfnamefont {A.~J.}\ \bibnamefont {Millis}},\ }\href
  {https://doi.org/10.1038/nphys883} {\bibfield  {journal} {\bibinfo  {journal}
  {Nature Physics}\ }\textbf {\bibinfo {volume} {4}},\ \bibinfo {pages} {287 EP
  } (\bibinfo {year} {2008})}\BibitemShut {NoStop}%
\bibitem [{\citenamefont {Ferrero}\ \emph {et~al.}(2010)\citenamefont
  {Ferrero}, \citenamefont {Parcollet}, \citenamefont {Georges}, \citenamefont
  {Kotliar},\ and\ \citenamefont {Basov}}]{Ferrero10}%
  \BibitemOpen
  \bibfield  {author} {\bibinfo {author} {\bibfnamefont {M.}~\bibnamefont
  {Ferrero}}, \bibinfo {author} {\bibfnamefont {O.}~\bibnamefont {Parcollet}},
  \bibinfo {author} {\bibfnamefont {A.}~\bibnamefont {Georges}}, \bibinfo
  {author} {\bibfnamefont {G.}~\bibnamefont {Kotliar}}, \ and\ \bibinfo
  {author} {\bibfnamefont {D.~N.}\ \bibnamefont {Basov}},\ }\href {\doibase
  10.1103/PhysRevB.82.054502} {\bibfield  {journal} {\bibinfo  {journal} {Phys.
  Rev. B}\ }\textbf {\bibinfo {volume} {82}},\ \bibinfo {pages} {054502}
  (\bibinfo {year} {2010})}\BibitemShut {NoStop}%
\bibitem [{\citenamefont {Bergeron}\ \emph {et~al.}(2011)\citenamefont
  {Bergeron}, \citenamefont {Hankevych}, \citenamefont {Kyung},\ and\
  \citenamefont {Tremblay}}]{Bergeron11}%
  \BibitemOpen
  \bibfield  {author} {\bibinfo {author} {\bibfnamefont {D.}~\bibnamefont
  {Bergeron}}, \bibinfo {author} {\bibfnamefont {V.}~\bibnamefont {Hankevych}},
  \bibinfo {author} {\bibfnamefont {B.}~\bibnamefont {Kyung}}, \ and\ \bibinfo
  {author} {\bibfnamefont {A.-M.~S.}\ \bibnamefont {Tremblay}},\ }\href
  {\doibase 10.1103/PhysRevB.84.085128} {\bibfield  {journal} {\bibinfo
  {journal} {Phys. Rev. B}\ }\textbf {\bibinfo {volume} {84}},\ \bibinfo
  {pages} {085128} (\bibinfo {year} {2011})}\BibitemShut {NoStop}%
\bibitem [{\citenamefont {Lin}\ \emph {et~al.}(2012)\citenamefont {Lin},
  \citenamefont {Gull},\ and\ \citenamefont {Millis}}]{Lin12}%
  \BibitemOpen
  \bibfield  {author} {\bibinfo {author} {\bibfnamefont {N.}~\bibnamefont
  {Lin}}, \bibinfo {author} {\bibfnamefont {E.}~\bibnamefont {Gull}}, \ and\
  \bibinfo {author} {\bibfnamefont {A.~J.}\ \bibnamefont {Millis}},\ }\href
  {\doibase 10.1103/PhysRevLett.109.106401} {\bibfield  {journal} {\bibinfo
  {journal} {Phys. Rev. Lett.}\ }\textbf {\bibinfo {volume} {109}},\ \bibinfo
  {pages} {106401} (\bibinfo {year} {2012})}\BibitemShut {NoStop}%
\bibitem [{\citenamefont {{Gull}}\ and\ \citenamefont
  {{Millis}}(2013)}]{Gull13_raman}%
  \BibitemOpen
  \bibfield  {author} {\bibinfo {author} {\bibfnamefont {E.}~\bibnamefont
  {{Gull}}}\ and\ \bibinfo {author} {\bibfnamefont {A.~J.}\ \bibnamefont
  {{Millis}}},\ }\href {\doibase 10.1103/PhysRevB.88.075127} {\bibfield
  {journal} {\bibinfo  {journal} {\prb}\ }\textbf {\bibinfo {volume} {88}},\
  \bibinfo {eid} {075127} (\bibinfo {year} {2013})},\ \Eprint
  {http://arxiv.org/abs/1304.6406} {1304.6406} \BibitemShut {NoStop}%
\bibitem [{\citenamefont {LeBlanc}\ \emph {et~al.}(2019)\citenamefont
  {LeBlanc}, \citenamefont {Li}, \citenamefont {Chen}, \citenamefont {Levy},
  \citenamefont {Antipov}, \citenamefont {Millis},\ and\ \citenamefont
  {Gull}}]{LeBlanc19_neutron}%
  \BibitemOpen
  \bibfield  {author} {\bibinfo {author} {\bibfnamefont {J.~P.~F.}\
  \bibnamefont {LeBlanc}}, \bibinfo {author} {\bibfnamefont {S.}~\bibnamefont
  {Li}}, \bibinfo {author} {\bibfnamefont {X.}~\bibnamefont {Chen}}, \bibinfo
  {author} {\bibfnamefont {R.}~\bibnamefont {Levy}}, \bibinfo {author}
  {\bibfnamefont {A.~E.}\ \bibnamefont {Antipov}}, \bibinfo {author}
  {\bibfnamefont {A.~J.}\ \bibnamefont {Millis}}, \ and\ \bibinfo {author}
  {\bibfnamefont {E.}~\bibnamefont {Gull}},\ }\href {\doibase
  10.1103/PhysRevB.100.075123} {\bibfield  {journal} {\bibinfo  {journal}
  {Phys. Rev. B}\ }\textbf {\bibinfo {volume} {100}},\ \bibinfo {pages}
  {075123} (\bibinfo {year} {2019})}\BibitemShut {NoStop}%
\bibitem [{\citenamefont {Macridin}\ \emph {et~al.}(2006)\citenamefont
  {Macridin}, \citenamefont {Jarrell}, \citenamefont {Maier}, \citenamefont
  {Kent},\ and\ \citenamefont {D'Azevedo}}]{Macridin06}%
  \BibitemOpen
  \bibfield  {author} {\bibinfo {author} {\bibfnamefont {A.}~\bibnamefont
  {Macridin}}, \bibinfo {author} {\bibfnamefont {M.}~\bibnamefont {Jarrell}},
  \bibinfo {author} {\bibfnamefont {T.}~\bibnamefont {Maier}}, \bibinfo
  {author} {\bibfnamefont {P.~R.~C.}\ \bibnamefont {Kent}}, \ and\ \bibinfo
  {author} {\bibfnamefont {E.}~\bibnamefont {D'Azevedo}},\ }\href {\doibase
  10.1103/PhysRevLett.97.036401} {\bibfield  {journal} {\bibinfo  {journal}
  {Phys. Rev. Lett.}\ }\textbf {\bibinfo {volume} {97}},\ \bibinfo {pages}
  {036401} (\bibinfo {year} {2006})}\BibitemShut {NoStop}%
\bibitem [{\citenamefont {Chen}\ \emph {et~al.}(2017)\citenamefont {Chen},
  \citenamefont {LeBlanc},\ and\ \citenamefont {Gull}}]{Chen17}%
  \BibitemOpen
  \bibfield  {author} {\bibinfo {author} {\bibfnamefont {X.}~\bibnamefont
  {Chen}}, \bibinfo {author} {\bibfnamefont {J.~P.~F.}\ \bibnamefont
  {LeBlanc}}, \ and\ \bibinfo {author} {\bibfnamefont {E.}~\bibnamefont
  {Gull}},\ }\href {https://doi.org/10.1038/ncomms14986} {\bibfield  {journal}
  {\bibinfo  {journal} {Nature Communications}\ }\textbf {\bibinfo {volume}
  {8}},\ \bibinfo {pages} {14986 EP } (\bibinfo {year} {2017})},\ \bibinfo
  {note} {article}\BibitemShut {NoStop}%
\bibitem [{\citenamefont {White}\ and\ \citenamefont
  {Scalapino}(2003)}]{White03_stripe}%
  \BibitemOpen
  \bibfield  {author} {\bibinfo {author} {\bibfnamefont {S.~R.}\ \bibnamefont
  {White}}\ and\ \bibinfo {author} {\bibfnamefont {D.~J.}\ \bibnamefont
  {Scalapino}},\ }\href {\doibase 10.1103/PhysRevLett.91.136403} {\bibfield
  {journal} {\bibinfo  {journal} {Phys. Rev. Lett.}\ }\textbf {\bibinfo
  {volume} {91}},\ \bibinfo {pages} {136403} (\bibinfo {year}
  {2003})}\BibitemShut {NoStop}%
\bibitem [{\citenamefont {Hager}\ \emph {et~al.}(2005)\citenamefont {Hager},
  \citenamefont {Wellein}, \citenamefont {Jeckelmann},\ and\ \citenamefont
  {Fehske}}]{Hager05_stripe}%
  \BibitemOpen
  \bibfield  {author} {\bibinfo {author} {\bibfnamefont {G.}~\bibnamefont
  {Hager}}, \bibinfo {author} {\bibfnamefont {G.}~\bibnamefont {Wellein}},
  \bibinfo {author} {\bibfnamefont {E.}~\bibnamefont {Jeckelmann}}, \ and\
  \bibinfo {author} {\bibfnamefont {H.}~\bibnamefont {Fehske}},\ }\href
  {\doibase 10.1103/PhysRevB.71.075108} {\bibfield  {journal} {\bibinfo
  {journal} {Phys. Rev. B}\ }\textbf {\bibinfo {volume} {71}},\ \bibinfo
  {pages} {075108} (\bibinfo {year} {2005})}\BibitemShut {NoStop}%
\bibitem [{\citenamefont {Chang}\ and\ \citenamefont
  {Zhang}(2010)}]{Chang08_strip}%
  \BibitemOpen
  \bibfield  {author} {\bibinfo {author} {\bibfnamefont {C.-C.}\ \bibnamefont
  {Chang}}\ and\ \bibinfo {author} {\bibfnamefont {S.}~\bibnamefont {Zhang}},\
  }\href {\doibase 10.1103/PhysRevLett.104.116402} {\bibfield  {journal}
  {\bibinfo  {journal} {Phys. Rev. Lett.}\ }\textbf {\bibinfo {volume} {104}},\
  \bibinfo {pages} {116402} (\bibinfo {year} {2010})}\BibitemShut {NoStop}%
\bibitem [{\citenamefont {Scalapino}(2012)}]{Scalapino12_stripe}%
  \BibitemOpen
  \bibfield  {author} {\bibinfo {author} {\bibfnamefont {D.~J.}\ \bibnamefont
  {Scalapino}},\ }\href {\doibase 10.1103/RevModPhys.84.1383} {\bibfield
  {journal} {\bibinfo  {journal} {Rev. Mod. Phys.}\ }\textbf {\bibinfo {volume}
  {84}},\ \bibinfo {pages} {1383} (\bibinfo {year} {2012})}\BibitemShut
  {NoStop}%
\bibitem [{\citenamefont {Zheng}\ and\ \citenamefont
  {Chan}(2016)}]{Zheng16_stripe}%
  \BibitemOpen
  \bibfield  {author} {\bibinfo {author} {\bibfnamefont {B.-X.}\ \bibnamefont
  {Zheng}}\ and\ \bibinfo {author} {\bibfnamefont {G.~K.-L.}\ \bibnamefont
  {Chan}},\ }\href {\doibase 10.1103/PhysRevB.93.035126} {\bibfield  {journal}
  {\bibinfo  {journal} {Phys. Rev. B}\ }\textbf {\bibinfo {volume} {93}},\
  \bibinfo {pages} {035126} (\bibinfo {year} {2016})}\BibitemShut {NoStop}%
\bibitem [{\citenamefont {Huang}\ \emph {et~al.}(2017)\citenamefont {Huang},
  \citenamefont {Mendl}, \citenamefont {Liu}, \citenamefont {Johnston},
  \citenamefont {Jiang}, \citenamefont {Moritz},\ and\ \citenamefont
  {Devereaux}}]{Huang17_stripe}%
  \BibitemOpen
  \bibfield  {author} {\bibinfo {author} {\bibfnamefont {E.~W.}\ \bibnamefont
  {Huang}}, \bibinfo {author} {\bibfnamefont {C.~B.}\ \bibnamefont {Mendl}},
  \bibinfo {author} {\bibfnamefont {S.}~\bibnamefont {Liu}}, \bibinfo {author}
  {\bibfnamefont {S.}~\bibnamefont {Johnston}}, \bibinfo {author}
  {\bibfnamefont {H.-C.}\ \bibnamefont {Jiang}}, \bibinfo {author}
  {\bibfnamefont {B.}~\bibnamefont {Moritz}}, \ and\ \bibinfo {author}
  {\bibfnamefont {T.~P.}\ \bibnamefont {Devereaux}},\ }\href {\doibase
  10.1126/science.aak9546} {\bibfield  {journal} {\bibinfo  {journal}
  {Science}\ }\textbf {\bibinfo {volume} {358}},\ \bibinfo {pages} {1161}
  (\bibinfo {year} {2017})}\BibitemShut {NoStop}%
\bibitem [{\citenamefont {Zheng}\ \emph {et~al.}(2017)\citenamefont {Zheng},
  \citenamefont {Chung}, \citenamefont {Corboz}, \citenamefont {Ehlers},
  \citenamefont {Qin}, \citenamefont {Noack}, \citenamefont {Shi},
  \citenamefont {White}, \citenamefont {Zhang},\ and\ \citenamefont
  {Chan}}]{Zheng17_stripe}%
  \BibitemOpen
  \bibfield  {author} {\bibinfo {author} {\bibfnamefont {B.-X.}\ \bibnamefont
  {Zheng}}, \bibinfo {author} {\bibfnamefont {C.-M.}\ \bibnamefont {Chung}},
  \bibinfo {author} {\bibfnamefont {P.}~\bibnamefont {Corboz}}, \bibinfo
  {author} {\bibfnamefont {G.}~\bibnamefont {Ehlers}}, \bibinfo {author}
  {\bibfnamefont {M.-P.}\ \bibnamefont {Qin}}, \bibinfo {author} {\bibfnamefont
  {R.~M.}\ \bibnamefont {Noack}}, \bibinfo {author} {\bibfnamefont
  {H.}~\bibnamefont {Shi}}, \bibinfo {author} {\bibfnamefont {S.~R.}\
  \bibnamefont {White}}, \bibinfo {author} {\bibfnamefont {S.}~\bibnamefont
  {Zhang}}, \ and\ \bibinfo {author} {\bibfnamefont {G.~K.-L.}\ \bibnamefont
  {Chan}},\ }\href {\doibase 10.1126/science.aam7127} {\bibfield  {journal}
  {\bibinfo  {journal} {Science}\ }\textbf {\bibinfo {volume} {358}},\ \bibinfo
  {pages} {1155} (\bibinfo {year} {2017})}\BibitemShut {NoStop}%
\bibitem [{\citenamefont {{Huang}}\ \emph {et~al.}(2018)\citenamefont
  {{Huang}}, \citenamefont {{Mendl}}, \citenamefont {{Jiang}}, \citenamefont
  {{Moritz}},\ and\ \citenamefont {{Devereaux}}}]{Huang18_stripe}%
  \BibitemOpen
  \bibfield  {author} {\bibinfo {author} {\bibfnamefont {E.~W.}\ \bibnamefont
  {{Huang}}}, \bibinfo {author} {\bibfnamefont {C.~B.}\ \bibnamefont
  {{Mendl}}}, \bibinfo {author} {\bibfnamefont {H.-C.}\ \bibnamefont
  {{Jiang}}}, \bibinfo {author} {\bibfnamefont {B.}~\bibnamefont {{Moritz}}}, \
  and\ \bibinfo {author} {\bibfnamefont {T.~P.}\ \bibnamefont {{Devereaux}}},\
  }\href {\doibase 10.1038/s41535-018-0097-0} {\bibfield  {journal} {\bibinfo
  {journal} {npj Quantum Materials}\ }\textbf {\bibinfo {volume} {3}},\
  \bibinfo {eid} {22} (\bibinfo {year} {2018})},\ \Eprint
  {http://arxiv.org/abs/1709.02398} {1709.02398} \BibitemShut {NoStop}%
\bibitem [{\citenamefont {Vanhala}\ and\ \citenamefont
  {T\"orm\"a}(2018)}]{Vanhala18_stripe}%
  \BibitemOpen
  \bibfield  {author} {\bibinfo {author} {\bibfnamefont {T.~I.}\ \bibnamefont
  {Vanhala}}\ and\ \bibinfo {author} {\bibfnamefont {P.}~\bibnamefont
  {T\"orm\"a}},\ }\href {\doibase 10.1103/PhysRevB.97.075112} {\bibfield
  {journal} {\bibinfo  {journal} {Phys. Rev. B}\ }\textbf {\bibinfo {volume}
  {97}},\ \bibinfo {pages} {075112} (\bibinfo {year} {2018})}\BibitemShut
  {NoStop}%
\bibitem [{\citenamefont {Darmawan}\ \emph {et~al.}(2018)\citenamefont
  {Darmawan}, \citenamefont {Nomura}, \citenamefont {Yamaji},\ and\
  \citenamefont {Imada}}]{Darmawan18_stripe}%
  \BibitemOpen
  \bibfield  {author} {\bibinfo {author} {\bibfnamefont {A.~S.}\ \bibnamefont
  {Darmawan}}, \bibinfo {author} {\bibfnamefont {Y.}~\bibnamefont {Nomura}},
  \bibinfo {author} {\bibfnamefont {Y.}~\bibnamefont {Yamaji}}, \ and\ \bibinfo
  {author} {\bibfnamefont {M.}~\bibnamefont {Imada}},\ }\href {\doibase
  10.1103/PhysRevB.98.205132} {\bibfield  {journal} {\bibinfo  {journal} {Phys.
  Rev. B}\ }\textbf {\bibinfo {volume} {98}},\ \bibinfo {pages} {205132}
  (\bibinfo {year} {2018})}\BibitemShut {NoStop}%
\bibitem [{\citenamefont {Tocchio}\ \emph {et~al.}(2019)\citenamefont
  {Tocchio}, \citenamefont {Montorsi},\ and\ \citenamefont
  {Becca}}]{Tocchio19_stripe}%
  \BibitemOpen
  \bibfield  {author} {\bibinfo {author} {\bibfnamefont {L.~F.}\ \bibnamefont
  {Tocchio}}, \bibinfo {author} {\bibfnamefont {A.}~\bibnamefont {Montorsi}}, \
  and\ \bibinfo {author} {\bibfnamefont {F.}~\bibnamefont {Becca}},\ }\href
  {\doibase 10.21468/SciPostPhys.7.2.021} {\bibfield  {journal} {\bibinfo
  {journal} {SciPost Phys.}\ }\textbf {\bibinfo {volume} {7}},\ \bibinfo
  {pages} {21} (\bibinfo {year} {2019})}\BibitemShut {NoStop}%
\bibitem [{\citenamefont {Zimmermann}\ \emph {et~al.}(1997)\citenamefont
  {Zimmermann}, \citenamefont {Fr\'esard},\ and\ \citenamefont
  {W\"olfle}}]{Zimmermann97}%
  \BibitemOpen
  \bibfield  {author} {\bibinfo {author} {\bibfnamefont {W.}~\bibnamefont
  {Zimmermann}}, \bibinfo {author} {\bibfnamefont {R.}~\bibnamefont
  {Fr\'esard}}, \ and\ \bibinfo {author} {\bibfnamefont {P.}~\bibnamefont
  {W\"olfle}},\ }\href {\doibase 10.1103/PhysRevB.56.10097} {\bibfield
  {journal} {\bibinfo  {journal} {Phys. Rev. B}\ }\textbf {\bibinfo {volume}
  {56}},\ \bibinfo {pages} {10097} (\bibinfo {year} {1997})}\BibitemShut
  {NoStop}%
\bibitem [{\citenamefont {Sherman}\ and\ \citenamefont
  {Schreiber}(2008)}]{Sherman08}%
  \BibitemOpen
  \bibfield  {author} {\bibinfo {author} {\bibfnamefont {A.}~\bibnamefont
  {Sherman}}\ and\ \bibinfo {author} {\bibfnamefont {M.}~\bibnamefont
  {Schreiber}},\ }\href {\doibase 10.1103/PhysRevB.77.155117} {\bibfield
  {journal} {\bibinfo  {journal} {Phys. Rev. B}\ }\textbf {\bibinfo {volume}
  {77}},\ \bibinfo {pages} {155117} (\bibinfo {year} {2008})}\BibitemShut
  {NoStop}%
\bibitem [{\citenamefont {Hochkeppel}\ \emph {et~al.}(2008)\citenamefont
  {Hochkeppel}, \citenamefont {Assaad},\ and\ \citenamefont
  {Hanke}}]{Hochkeppel08}%
  \BibitemOpen
  \bibfield  {author} {\bibinfo {author} {\bibfnamefont {S.}~\bibnamefont
  {Hochkeppel}}, \bibinfo {author} {\bibfnamefont {F.~F.}\ \bibnamefont
  {Assaad}}, \ and\ \bibinfo {author} {\bibfnamefont {W.}~\bibnamefont
  {Hanke}},\ }\href {\doibase 10.1103/PhysRevB.77.205103} {\bibfield  {journal}
  {\bibinfo  {journal} {Phys. Rev. B}\ }\textbf {\bibinfo {volume} {77}},\
  \bibinfo {pages} {205103} (\bibinfo {year} {2008})}\BibitemShut {NoStop}%
\bibitem [{\citenamefont {Kung}\ \emph {et~al.}(2015)\citenamefont {Kung},
  \citenamefont {Nowadnick}, \citenamefont {Jia}, \citenamefont {Johnston},
  \citenamefont {Moritz}, \citenamefont {Scalettar},\ and\ \citenamefont
  {Devereaux}}]{Devereaux15}%
  \BibitemOpen
  \bibfield  {author} {\bibinfo {author} {\bibfnamefont {Y.~F.}\ \bibnamefont
  {Kung}}, \bibinfo {author} {\bibfnamefont {E.~A.}\ \bibnamefont {Nowadnick}},
  \bibinfo {author} {\bibfnamefont {C.~J.}\ \bibnamefont {Jia}}, \bibinfo
  {author} {\bibfnamefont {S.}~\bibnamefont {Johnston}}, \bibinfo {author}
  {\bibfnamefont {B.}~\bibnamefont {Moritz}}, \bibinfo {author} {\bibfnamefont
  {R.~T.}\ \bibnamefont {Scalettar}}, \ and\ \bibinfo {author} {\bibfnamefont
  {T.~P.}\ \bibnamefont {Devereaux}},\ }\href {\doibase
  10.1103/PhysRevB.92.195108} {\bibfield  {journal} {\bibinfo  {journal} {Phys.
  Rev. B}\ }\textbf {\bibinfo {volume} {92}},\ \bibinfo {pages} {195108}
  (\bibinfo {year} {2015})}\BibitemShut {NoStop}%
\bibitem [{\citenamefont {{Lin}}\ \emph {et~al.}(2019)\citenamefont {{Lin}},
  \citenamefont {{Yuan}}, \citenamefont {{Jin}}, \citenamefont {{Yin}},
  \citenamefont {{Li}}, \citenamefont {{Zhou}}, \citenamefont {{Lu}},
  \citenamefont {{Dantz}}, \citenamefont {{Schmitt}}, \citenamefont {{Ding}},
  \citenamefont {{Guo}}, \citenamefont {{Dean}},\ and\ \citenamefont
  {{Liu}}}]{Lin19}%
  \BibitemOpen
  \bibfield  {author} {\bibinfo {author} {\bibfnamefont {J.~Q.}\ \bibnamefont
  {{Lin}}}, \bibinfo {author} {\bibfnamefont {J.}~\bibnamefont {{Yuan}}},
  \bibinfo {author} {\bibfnamefont {K.}~\bibnamefont {{Jin}}}, \bibinfo
  {author} {\bibfnamefont {Z.~P.}\ \bibnamefont {{Yin}}}, \bibinfo {author}
  {\bibfnamefont {G.}~\bibnamefont {{Li}}}, \bibinfo {author} {\bibfnamefont
  {K.-J.}\ \bibnamefont {{Zhou}}}, \bibinfo {author} {\bibfnamefont
  {X.}~\bibnamefont {{Lu}}}, \bibinfo {author} {\bibfnamefont {M.}~\bibnamefont
  {{Dantz}}}, \bibinfo {author} {\bibfnamefont {T.}~\bibnamefont {{Schmitt}}},
  \bibinfo {author} {\bibfnamefont {H.}~\bibnamefont {{Ding}}}, \bibinfo
  {author} {\bibfnamefont {H.}~\bibnamefont {{Guo}}}, \bibinfo {author}
  {\bibfnamefont {M.~P.~M.}\ \bibnamefont {{Dean}}}, \ and\ \bibinfo {author}
  {\bibfnamefont {X.}~\bibnamefont {{Liu}}},\ }\href@noop {} {\enquote
  {\bibinfo {title} {{Doping evolution of the charge excitations and electron
  correlations in electron-doped superconducting
  La$_{2-x}$Ce$_{x}$CuO$_{4}$}},}\ } (\bibinfo {year} {2019}),\ \Eprint
  {http://arxiv.org/abs/1906.11354} {arXiv:1906.11354} \BibitemShut {NoStop}%
\bibitem [{\citenamefont {Maier}\ \emph
  {et~al.}(2005{\natexlab{a}})\citenamefont {Maier}, \citenamefont {Jarrell},
  \citenamefont {Pruschke},\ and\ \citenamefont {Hettler}}]{Maier05}%
  \BibitemOpen
  \bibfield  {author} {\bibinfo {author} {\bibfnamefont {T.~A.}\ \bibnamefont
  {Maier}}, \bibinfo {author} {\bibfnamefont {M.}~\bibnamefont {Jarrell}},
  \bibinfo {author} {\bibfnamefont {T.}~\bibnamefont {Pruschke}}, \ and\
  \bibinfo {author} {\bibfnamefont {M.}~\bibnamefont {Hettler}},\ }\href
  {\doibase 10.1103/RevModPhys.77.1027} {\bibfield  {journal} {\bibinfo
  {journal} {Rev. Mod. Phys.}\ }\textbf {\bibinfo {volume} {77}},\ \bibinfo
  {eid} {1027} (\bibinfo {year} {2005}{\natexlab{a}})}\BibitemShut {NoStop}%
\bibitem [{\citenamefont {Andersen}\ \emph {et~al.}(1994)\citenamefont
  {Andersen}, \citenamefont {Jepsen}, \citenamefont {Liechtenstein},\ and\
  \citenamefont {Mazin}}]{Anderson94}%
  \BibitemOpen
  \bibfield  {author} {\bibinfo {author} {\bibfnamefont {O.~K.}\ \bibnamefont
  {Andersen}}, \bibinfo {author} {\bibfnamefont {O.}~\bibnamefont {Jepsen}},
  \bibinfo {author} {\bibfnamefont {A.~I.}\ \bibnamefont {Liechtenstein}}, \
  and\ \bibinfo {author} {\bibfnamefont {I.~I.}\ \bibnamefont {Mazin}},\ }\href
  {\doibase 10.1103/PhysRevB.49.4145} {\bibfield  {journal} {\bibinfo
  {journal} {Phys. Rev. B}\ }\textbf {\bibinfo {volume} {49}},\ \bibinfo
  {pages} {4145} (\bibinfo {year} {1994})}\BibitemShut {NoStop}%
\bibitem [{\citenamefont {Andersen}\ \emph {et~al.}(1995)\citenamefont
  {Andersen}, \citenamefont {Liechtenstein}, \citenamefont {Jepsen},\ and\
  \citenamefont {Paulsen}}]{Anderson95}%
  \BibitemOpen
  \bibfield  {author} {\bibinfo {author} {\bibfnamefont {O.}~\bibnamefont
  {Andersen}}, \bibinfo {author} {\bibfnamefont {A.}~\bibnamefont
  {Liechtenstein}}, \bibinfo {author} {\bibfnamefont {O.}~\bibnamefont
  {Jepsen}}, \ and\ \bibinfo {author} {\bibfnamefont {F.}~\bibnamefont
  {Paulsen}},\ }\href {\doibase https://doi.org/10.1016/0022-3697(95)00269-3}
  {\bibfield  {journal} {\bibinfo  {journal} {Journal of Physics and Chemistry
  of Solids}\ }\textbf {\bibinfo {volume} {56}},\ \bibinfo {pages} {1573 }
  (\bibinfo {year} {1995})},\ \bibinfo {note} {proceedings of the Conference on
  Spectroscopies in Novel Superconductors}\BibitemShut {NoStop}%
\bibitem [{\citenamefont {Kim}\ \emph {et~al.}(1998)\citenamefont {Kim},
  \citenamefont {White}, \citenamefont {Shen}, \citenamefont {Tohyama},
  \citenamefont {Shibata}, \citenamefont {Maekawa}, \citenamefont {Wells},
  \citenamefont {Kim}, \citenamefont {Birgeneau},\ and\ \citenamefont
  {Kastner}}]{Kim98_t}%
  \BibitemOpen
  \bibfield  {author} {\bibinfo {author} {\bibfnamefont {C.}~\bibnamefont
  {Kim}}, \bibinfo {author} {\bibfnamefont {P.~J.}\ \bibnamefont {White}},
  \bibinfo {author} {\bibfnamefont {Z.-X.}\ \bibnamefont {Shen}}, \bibinfo
  {author} {\bibfnamefont {T.}~\bibnamefont {Tohyama}}, \bibinfo {author}
  {\bibfnamefont {Y.}~\bibnamefont {Shibata}}, \bibinfo {author} {\bibfnamefont
  {S.}~\bibnamefont {Maekawa}}, \bibinfo {author} {\bibfnamefont {B.~O.}\
  \bibnamefont {Wells}}, \bibinfo {author} {\bibfnamefont {Y.~J.}\ \bibnamefont
  {Kim}}, \bibinfo {author} {\bibfnamefont {R.~J.}\ \bibnamefont {Birgeneau}},
  \ and\ \bibinfo {author} {\bibfnamefont {M.~A.}\ \bibnamefont {Kastner}},\
  }\href {\doibase 10.1103/PhysRevLett.80.4245} {\bibfield  {journal} {\bibinfo
   {journal} {Phys. Rev. Lett.}\ }\textbf {\bibinfo {volume} {80}},\ \bibinfo
  {pages} {4245} (\bibinfo {year} {1998})}\BibitemShut {NoStop}%
\bibitem [{\citenamefont {Hasan}\ \emph {et~al.}(2000)\citenamefont {Hasan},
  \citenamefont {Isaacs}, \citenamefont {Shen}, \citenamefont {Miller},
  \citenamefont {Tsutsui}, \citenamefont {Tohyama},\ and\ \citenamefont
  {Maekawa}}]{Hasan00}%
  \BibitemOpen
  \bibfield  {author} {\bibinfo {author} {\bibfnamefont {M.~Z.}\ \bibnamefont
  {Hasan}}, \bibinfo {author} {\bibfnamefont {E.~D.}\ \bibnamefont {Isaacs}},
  \bibinfo {author} {\bibfnamefont {Z.-X.}\ \bibnamefont {Shen}}, \bibinfo
  {author} {\bibfnamefont {L.~L.}\ \bibnamefont {Miller}}, \bibinfo {author}
  {\bibfnamefont {K.}~\bibnamefont {Tsutsui}}, \bibinfo {author} {\bibfnamefont
  {T.}~\bibnamefont {Tohyama}}, \ and\ \bibinfo {author} {\bibfnamefont
  {S.}~\bibnamefont {Maekawa}},\ }\href {\doibase
  10.1126/science.288.5472.1811} {\bibfield  {journal} {\bibinfo  {journal}
  {Science}\ }\textbf {\bibinfo {volume} {288}},\ \bibinfo {pages} {1811}
  (\bibinfo {year} {2000})}\BibitemShut {NoStop}%
\bibitem [{\citenamefont {Gull}\ \emph {et~al.}(2009)\citenamefont {Gull},
  \citenamefont {Parcollet}, \citenamefont {Werner},\ and\ \citenamefont
  {Millis}}]{Gull09_8site}%
  \BibitemOpen
  \bibfield  {author} {\bibinfo {author} {\bibfnamefont {E.}~\bibnamefont
  {Gull}}, \bibinfo {author} {\bibfnamefont {O.}~\bibnamefont {Parcollet}},
  \bibinfo {author} {\bibfnamefont {P.}~\bibnamefont {Werner}}, \ and\ \bibinfo
  {author} {\bibfnamefont {A.~J.}\ \bibnamefont {Millis}},\ }\href {\doibase
  10.1103/PhysRevB.80.245102} {\bibfield  {journal} {\bibinfo  {journal} {Phys.
  Rev. B}\ }\textbf {\bibinfo {volume} {80}},\ \bibinfo {pages} {245102}
  (\bibinfo {year} {2009})}\BibitemShut {NoStop}%
\bibitem [{\citenamefont {Gull}\ and\ \citenamefont
  {Millis}(2012)}]{Gull12_energy}%
  \BibitemOpen
  \bibfield  {author} {\bibinfo {author} {\bibfnamefont {E.}~\bibnamefont
  {Gull}}\ and\ \bibinfo {author} {\bibfnamefont {A.~J.}\ \bibnamefont
  {Millis}},\ }\href {\doibase 10.1103/PhysRevB.86.241106} {\bibfield
  {journal} {\bibinfo  {journal} {Phys. Rev. B}\ }\textbf {\bibinfo {volume}
  {86}},\ \bibinfo {pages} {241106(R)} (\bibinfo {year} {2012})}\BibitemShut
  {NoStop}%
\bibitem [{\citenamefont {Gull}\ \emph {et~al.}(2013)\citenamefont {Gull},
  \citenamefont {Parcollet},\ and\ \citenamefont {Millis}}]{Gull13_super}%
  \BibitemOpen
  \bibfield  {author} {\bibinfo {author} {\bibfnamefont {E.}~\bibnamefont
  {Gull}}, \bibinfo {author} {\bibfnamefont {O.}~\bibnamefont {Parcollet}}, \
  and\ \bibinfo {author} {\bibfnamefont {A.~J.}\ \bibnamefont {Millis}},\
  }\href {\doibase 10.1103/PhysRevLett.110.216405} {\bibfield  {journal}
  {\bibinfo  {journal} {Phys. Rev. Lett.}\ }\textbf {\bibinfo {volume} {110}},\
  \bibinfo {pages} {216405} (\bibinfo {year} {2013})}\BibitemShut {NoStop}%
\bibitem [{\citenamefont {Gull}\ and\ \citenamefont
  {Millis}(2014)}]{Gull14_pairing}%
  \BibitemOpen
  \bibfield  {author} {\bibinfo {author} {\bibfnamefont {E.}~\bibnamefont
  {Gull}}\ and\ \bibinfo {author} {\bibfnamefont {A.~J.}\ \bibnamefont
  {Millis}},\ }\href {\doibase 10.1103/PhysRevB.90.041110} {\bibfield
  {journal} {\bibinfo  {journal} {Phys. Rev. B}\ }\textbf {\bibinfo {volume}
  {90}},\ \bibinfo {pages} {041110(R)} (\bibinfo {year} {2014})}\BibitemShut
  {NoStop}%
\bibitem [{\citenamefont {Gull}\ and\ \citenamefont
  {Millis}(2015)}]{Gull15_qp}%
  \BibitemOpen
  \bibfield  {author} {\bibinfo {author} {\bibfnamefont {E.}~\bibnamefont
  {Gull}}\ and\ \bibinfo {author} {\bibfnamefont {A.~J.}\ \bibnamefont
  {Millis}},\ }\href {\doibase 10.1103/PhysRevB.91.085116} {\bibfield
  {journal} {\bibinfo  {journal} {Phys. Rev. B}\ }\textbf {\bibinfo {volume}
  {91}},\ \bibinfo {pages} {085116} (\bibinfo {year} {2015})}\BibitemShut
  {NoStop}%
\bibitem [{\citenamefont {Gull}\ \emph {et~al.}(2010)\citenamefont {Gull},
  \citenamefont {Ferrero}, \citenamefont {Parcollet}, \citenamefont {Georges},\
  and\ \citenamefont {Millis}}]{Gull10_clustercompare}%
  \BibitemOpen
  \bibfield  {author} {\bibinfo {author} {\bibfnamefont {E.}~\bibnamefont
  {Gull}}, \bibinfo {author} {\bibfnamefont {M.}~\bibnamefont {Ferrero}},
  \bibinfo {author} {\bibfnamefont {O.}~\bibnamefont {Parcollet}}, \bibinfo
  {author} {\bibfnamefont {A.}~\bibnamefont {Georges}}, \ and\ \bibinfo
  {author} {\bibfnamefont {A.~J.}\ \bibnamefont {Millis}},\ }\href {\doibase
  10.1103/PhysRevB.82.155101} {\bibfield  {journal} {\bibinfo  {journal} {Phys.
  Rev. B}\ }\textbf {\bibinfo {volume} {82}},\ \bibinfo {pages} {155101}
  (\bibinfo {year} {2010})}\BibitemShut {NoStop}%
\bibitem [{\citenamefont {Fuhrmann}\ \emph {et~al.}(2007)\citenamefont
  {Fuhrmann}, \citenamefont {Okamoto}, \citenamefont {Monien},\ and\
  \citenamefont {Millis}}]{Fuhrmann07}%
  \BibitemOpen
  \bibfield  {author} {\bibinfo {author} {\bibfnamefont {A.}~\bibnamefont
  {Fuhrmann}}, \bibinfo {author} {\bibfnamefont {S.}~\bibnamefont {Okamoto}},
  \bibinfo {author} {\bibfnamefont {H.}~\bibnamefont {Monien}}, \ and\ \bibinfo
  {author} {\bibfnamefont {A.~J.}\ \bibnamefont {Millis}},\ }\href {\doibase
  10.1103/PhysRevB.75.205118} {\bibfield  {journal} {\bibinfo  {journal} {Phys.
  Rev. B}\ }\textbf {\bibinfo {volume} {75}},\ \bibinfo {pages} {205118}
  (\bibinfo {year} {2007})}\BibitemShut {NoStop}%
\bibitem [{\citenamefont {Maier}\ \emph
  {et~al.}(2005{\natexlab{b}})\citenamefont {Maier}, \citenamefont {Jarrell},
  \citenamefont {Schulthess}, \citenamefont {Kent},\ and\ \citenamefont
  {White}}]{Maier05_dwave}%
  \BibitemOpen
  \bibfield  {author} {\bibinfo {author} {\bibfnamefont {T.~A.}\ \bibnamefont
  {Maier}}, \bibinfo {author} {\bibfnamefont {M.}~\bibnamefont {Jarrell}},
  \bibinfo {author} {\bibfnamefont {T.~C.}\ \bibnamefont {Schulthess}},
  \bibinfo {author} {\bibfnamefont {P.~R.~C.}\ \bibnamefont {Kent}}, \ and\
  \bibinfo {author} {\bibfnamefont {J.~B.}\ \bibnamefont {White}},\ }\href
  {\doibase 10.1103/PhysRevLett.95.237001} {\bibfield  {journal} {\bibinfo
  {journal} {Phys. Rev. Lett.}\ }\textbf {\bibinfo {volume} {95}},\ \bibinfo
  {pages} {237001} (\bibinfo {year} {2005}{\natexlab{b}})}\BibitemShut
  {NoStop}%
\bibitem [{\citenamefont {Fuchs}\ \emph {et~al.}(2011)\citenamefont {Fuchs},
  \citenamefont {Gull}, \citenamefont {Pollet}, \citenamefont {Burovski},
  \citenamefont {Kozik}, \citenamefont {Pruschke},\ and\ \citenamefont
  {Troyer}}]{Fuchs11}%
  \BibitemOpen
  \bibfield  {author} {\bibinfo {author} {\bibfnamefont {S.}~\bibnamefont
  {Fuchs}}, \bibinfo {author} {\bibfnamefont {E.}~\bibnamefont {Gull}},
  \bibinfo {author} {\bibfnamefont {L.}~\bibnamefont {Pollet}}, \bibinfo
  {author} {\bibfnamefont {E.}~\bibnamefont {Burovski}}, \bibinfo {author}
  {\bibfnamefont {E.}~\bibnamefont {Kozik}}, \bibinfo {author} {\bibfnamefont
  {T.}~\bibnamefont {Pruschke}}, \ and\ \bibinfo {author} {\bibfnamefont
  {M.}~\bibnamefont {Troyer}},\ }\href@noop {} {\bibfield  {journal} {\bibinfo
  {journal} {Phys. Rev. Lett.}\ }\textbf {\bibinfo {volume} {106}},\ \bibinfo
  {pages} {030401} (\bibinfo {year} {2011})}\BibitemShut {NoStop}%
\bibitem [{\citenamefont {LeBlanc}\ \emph {et~al.}(2015)\citenamefont
  {LeBlanc}, \citenamefont {Antipov}, \citenamefont {Becca}, \citenamefont
  {Bulik}, \citenamefont {Chan}, \citenamefont {Chung}, \citenamefont {Deng},
  \citenamefont {Ferrero}, \citenamefont {Henderson}, \citenamefont
  {Jim\'enez-Hoyos}, \citenamefont {Kozik}, \citenamefont {Liu}, \citenamefont
  {Millis}, \citenamefont {Prokof'ev}, \citenamefont {Qin}, \citenamefont
  {Scuseria}, \citenamefont {Shi}, \citenamefont {Svistunov}, \citenamefont
  {Tocchio}, \citenamefont {Tupitsyn}, \citenamefont {White}, \citenamefont
  {Zhang}, \citenamefont {Zheng}, \citenamefont {Zhu},\ and\ \citenamefont
  {Gull}}]{LeBlanc15}%
  \BibitemOpen
  \bibfield  {author} {\bibinfo {author} {\bibfnamefont {J.~P.~F.}\
  \bibnamefont {LeBlanc}}, \bibinfo {author} {\bibfnamefont {A.~E.}\
  \bibnamefont {Antipov}}, \bibinfo {author} {\bibfnamefont {F.}~\bibnamefont
  {Becca}}, \bibinfo {author} {\bibfnamefont {I.~W.}\ \bibnamefont {Bulik}},
  \bibinfo {author} {\bibfnamefont {G.~K.-L.}\ \bibnamefont {Chan}}, \bibinfo
  {author} {\bibfnamefont {C.-M.}\ \bibnamefont {Chung}}, \bibinfo {author}
  {\bibfnamefont {Y.}~\bibnamefont {Deng}}, \bibinfo {author} {\bibfnamefont
  {M.}~\bibnamefont {Ferrero}}, \bibinfo {author} {\bibfnamefont {T.~M.}\
  \bibnamefont {Henderson}}, \bibinfo {author} {\bibfnamefont {C.~A.}\
  \bibnamefont {Jim\'enez-Hoyos}}, \bibinfo {author} {\bibfnamefont
  {E.}~\bibnamefont {Kozik}}, \bibinfo {author} {\bibfnamefont {X.-W.}\
  \bibnamefont {Liu}}, \bibinfo {author} {\bibfnamefont {A.~J.}\ \bibnamefont
  {Millis}}, \bibinfo {author} {\bibfnamefont {N.~V.}\ \bibnamefont
  {Prokof'ev}}, \bibinfo {author} {\bibfnamefont {M.}~\bibnamefont {Qin}},
  \bibinfo {author} {\bibfnamefont {G.~E.}\ \bibnamefont {Scuseria}}, \bibinfo
  {author} {\bibfnamefont {H.}~\bibnamefont {Shi}}, \bibinfo {author}
  {\bibfnamefont {B.~V.}\ \bibnamefont {Svistunov}}, \bibinfo {author}
  {\bibfnamefont {L.~F.}\ \bibnamefont {Tocchio}}, \bibinfo {author}
  {\bibfnamefont {I.~S.}\ \bibnamefont {Tupitsyn}}, \bibinfo {author}
  {\bibfnamefont {S.~R.}\ \bibnamefont {White}}, \bibinfo {author}
  {\bibfnamefont {S.}~\bibnamefont {Zhang}}, \bibinfo {author} {\bibfnamefont
  {B.-X.}\ \bibnamefont {Zheng}}, \bibinfo {author} {\bibfnamefont
  {Z.}~\bibnamefont {Zhu}}, \ and\ \bibinfo {author} {\bibfnamefont
  {E.}~\bibnamefont {Gull}} (\bibinfo {collaboration} {Simons Collaboration on
  the Many-Electron Problem}),\ }\href {\doibase 10.1103/PhysRevX.5.041041}
  {\bibfield  {journal} {\bibinfo  {journal} {Phys. Rev. X}\ }\textbf {\bibinfo
  {volume} {5}},\ \bibinfo {pages} {041041} (\bibinfo {year}
  {2015})}\BibitemShut {NoStop}%
\bibitem [{\citenamefont {Gull}\ \emph {et~al.}(2008)\citenamefont {Gull},
  \citenamefont {Werner}, \citenamefont {Parcollet},\ and\ \citenamefont
  {Troyer}}]{Gull08}%
  \BibitemOpen
  \bibfield  {author} {\bibinfo {author} {\bibfnamefont {E.}~\bibnamefont
  {Gull}}, \bibinfo {author} {\bibfnamefont {P.}~\bibnamefont {Werner}},
  \bibinfo {author} {\bibfnamefont {O.}~\bibnamefont {Parcollet}}, \ and\
  \bibinfo {author} {\bibfnamefont {M.}~\bibnamefont {Troyer}},\ }\href
  {\doibase 10.1209/0295-5075/82/57003} {\bibfield  {journal} {\bibinfo
  {journal} {Europhys. Lett.}\ }\textbf {\bibinfo {volume} {82}},\ \bibinfo
  {pages} {57003} (\bibinfo {year} {2008})}\BibitemShut {NoStop}%
\bibitem [{\citenamefont {Gull}\ \emph
  {et~al.}(2011{\natexlab{a}})\citenamefont {Gull}, \citenamefont {Millis},
  \citenamefont {Lichtenstein}, \citenamefont {Rubtsov}, \citenamefont
  {Troyer},\ and\ \citenamefont {Werner}}]{Gull11}%
  \BibitemOpen
  \bibfield  {author} {\bibinfo {author} {\bibfnamefont {E.}~\bibnamefont
  {Gull}}, \bibinfo {author} {\bibfnamefont {A.~J.}\ \bibnamefont {Millis}},
  \bibinfo {author} {\bibfnamefont {A.~I.}\ \bibnamefont {Lichtenstein}},
  \bibinfo {author} {\bibfnamefont {A.~N.}\ \bibnamefont {Rubtsov}}, \bibinfo
  {author} {\bibfnamefont {M.}~\bibnamefont {Troyer}}, \ and\ \bibinfo {author}
  {\bibfnamefont {P.}~\bibnamefont {Werner}},\ }\href {\doibase
  10.1103/RevModPhys.83.349} {\bibfield  {journal} {\bibinfo  {journal} {Rev.
  Mod. Phys.}\ }\textbf {\bibinfo {volume} {83}},\ \bibinfo {pages} {349}
  (\bibinfo {year} {2011}{\natexlab{a}})}\BibitemShut {NoStop}%
\bibitem [{\citenamefont {Gull}\ \emph
  {et~al.}(2011{\natexlab{b}})\citenamefont {Gull}, \citenamefont {Staar},
  \citenamefont {Fuchs}, \citenamefont {Nukala}, \citenamefont {Summers},
  \citenamefont {Pruschke}, \citenamefont {Schulthess},\ and\ \citenamefont
  {Maier}}]{Gull10_submatrix}%
  \BibitemOpen
  \bibfield  {author} {\bibinfo {author} {\bibfnamefont {E.}~\bibnamefont
  {Gull}}, \bibinfo {author} {\bibfnamefont {P.}~\bibnamefont {Staar}},
  \bibinfo {author} {\bibfnamefont {S.}~\bibnamefont {Fuchs}}, \bibinfo
  {author} {\bibfnamefont {P.}~\bibnamefont {Nukala}}, \bibinfo {author}
  {\bibfnamefont {M.~S.}\ \bibnamefont {Summers}}, \bibinfo {author}
  {\bibfnamefont {T.}~\bibnamefont {Pruschke}}, \bibinfo {author}
  {\bibfnamefont {T.~C.}\ \bibnamefont {Schulthess}}, \ and\ \bibinfo {author}
  {\bibfnamefont {T.}~\bibnamefont {Maier}},\ }\href {\doibase
  10.1103/PhysRevB.83.075122} {\bibfield  {journal} {\bibinfo  {journal} {Phys.
  Rev. B}\ }\textbf {\bibinfo {volume} {83}},\ \bibinfo {pages} {075122}
  (\bibinfo {year} {2011}{\natexlab{b}})}\BibitemShut {NoStop}%
\bibitem [{\citenamefont {Rohringer}\ \emph {et~al.}(2012)\citenamefont
  {Rohringer}, \citenamefont {Valli},\ and\ \citenamefont
  {Toschi}}]{Roheringer12}%
  \BibitemOpen
  \bibfield  {author} {\bibinfo {author} {\bibfnamefont {G.}~\bibnamefont
  {Rohringer}}, \bibinfo {author} {\bibfnamefont {A.}~\bibnamefont {Valli}}, \
  and\ \bibinfo {author} {\bibfnamefont {A.}~\bibnamefont {Toschi}},\ }\href
  {\doibase 10.1103/PhysRevB.86.125114} {\bibfield  {journal} {\bibinfo
  {journal} {Phys. Rev. B}\ }\textbf {\bibinfo {volume} {86}},\ \bibinfo
  {pages} {125114} (\bibinfo {year} {2012})}\BibitemShut {NoStop}%
\bibitem [{\citenamefont {Jarrell}\ and\ \citenamefont
  {Gubernatis}(1996)}]{Jarrell96}%
  \BibitemOpen
  \bibfield  {author} {\bibinfo {author} {\bibfnamefont {M.}~\bibnamefont
  {Jarrell}}\ and\ \bibinfo {author} {\bibfnamefont {J.~E.}\ \bibnamefont
  {Gubernatis}},\ }\href
  {http://www.sciencedirect.com/science/article/B6TVP-3VTNGGS-3/2/5f7d5c3a6ee0870df8970b51e47592d2}
  {\bibfield  {journal} {\bibinfo  {journal} {Physics Reports}\ }\textbf
  {\bibinfo {volume} {269}},\ \bibinfo {pages} {133} (\bibinfo {year}
  {1996})}\BibitemShut {NoStop}%
\bibitem [{\citenamefont {{Levy}}\ \emph {et~al.}(2017)\citenamefont {{Levy}},
  \citenamefont {{LeBlanc}},\ and\ \citenamefont {{Gull}}}]{Ryan17}%
  \BibitemOpen
  \bibfield  {author} {\bibinfo {author} {\bibfnamefont {R.}~\bibnamefont
  {{Levy}}}, \bibinfo {author} {\bibfnamefont {J.~P.~F.}\ \bibnamefont
  {{LeBlanc}}}, \ and\ \bibinfo {author} {\bibfnamefont {E.}~\bibnamefont
  {{Gull}}},\ }\href {\doibase 10.1016/j.cpc.2017.01.018} {\bibfield  {journal}
  {\bibinfo  {journal} {Computer Physics Communications}\ }\textbf {\bibinfo
  {volume} {215}},\ \bibinfo {pages} {149} (\bibinfo {year} {2017})},\ \Eprint
  {http://arxiv.org/abs/1606.00368} {1606.00368} \BibitemShut {NoStop}%
\bibitem [{\citenamefont {Hafermann}\ \emph {et~al.}(2014)\citenamefont
  {Hafermann}, \citenamefont {van Loon}, \citenamefont {Katsnelson},
  \citenamefont {Lichtenstein},\ and\ \citenamefont
  {Parcollet}}]{Ward_Identity}%
  \BibitemOpen
  \bibfield  {author} {\bibinfo {author} {\bibfnamefont {H.}~\bibnamefont
  {Hafermann}}, \bibinfo {author} {\bibfnamefont {E.~G. C.~P.}\ \bibnamefont
  {van Loon}}, \bibinfo {author} {\bibfnamefont {M.~I.}\ \bibnamefont
  {Katsnelson}}, \bibinfo {author} {\bibfnamefont {A.~I.}\ \bibnamefont
  {Lichtenstein}}, \ and\ \bibinfo {author} {\bibfnamefont {O.}~\bibnamefont
  {Parcollet}},\ }\href {\doibase 10.1103/PhysRevB.90.235105} {\bibfield
  {journal} {\bibinfo  {journal} {Phys. Rev. B}\ }\textbf {\bibinfo {volume}
  {90}},\ \bibinfo {pages} {235105} (\bibinfo {year} {2014})}\BibitemShut
  {NoStop}%
\bibitem [{\citenamefont {Bohm}\ and\ \citenamefont {Pines}(1951)}]{Bohm51}%
  \BibitemOpen
  \bibfield  {author} {\bibinfo {author} {\bibfnamefont {D.}~\bibnamefont
  {Bohm}}\ and\ \bibinfo {author} {\bibfnamefont {D.}~\bibnamefont {Pines}},\
  }\href {\doibase 10.1103/PhysRev.82.625} {\bibfield  {journal} {\bibinfo
  {journal} {Phys. Rev.}\ }\textbf {\bibinfo {volume} {82}},\ \bibinfo {pages}
  {625} (\bibinfo {year} {1951})}\BibitemShut {NoStop}%
\bibitem [{\citenamefont {Hedin}(1965)}]{Hedin64}%
  \BibitemOpen
  \bibfield  {author} {\bibinfo {author} {\bibfnamefont {L.}~\bibnamefont
  {Hedin}},\ }\href {\doibase 10.1103/PhysRev.139.A796} {\bibfield  {journal}
  {\bibinfo  {journal} {Phys. Rev.}\ }\textbf {\bibinfo {volume} {139}},\
  \bibinfo {pages} {A796} (\bibinfo {year} {1965})}\BibitemShut {NoStop}%
\bibitem [{\citenamefont {Bickers}(2004)}]{Bickers2004}%
  \BibitemOpen
  \bibfield  {author} {\bibinfo {author} {\bibfnamefont {N.~E.}\ \bibnamefont
  {Bickers}},\ }\enquote {\bibinfo {title} {Self-consistent many-body theory
  for condensed matter systems},}\ in\ \href {\doibase 10.1007/0-387-21717-7_6}
  {\emph {\bibinfo {booktitle} {Theoretical Methods for Strongly Correlated
  Electrons}}}\ (\bibinfo  {publisher} {Springer New York},\ \bibinfo {address}
  {New York, NY},\ \bibinfo {year} {2004})\ pp.\ \bibinfo {pages}
  {237--296}\BibitemShut {NoStop}%
\bibitem [{\citenamefont {Brener}\ \emph {et~al.}(2008)\citenamefont {Brener},
  \citenamefont {Hafermann}, \citenamefont {Rubtsov}, \citenamefont
  {Katsnelson},\ and\ \citenamefont {Lichtenstein}}]{Brener08}%
  \BibitemOpen
  \bibfield  {author} {\bibinfo {author} {\bibfnamefont {S.}~\bibnamefont
  {Brener}}, \bibinfo {author} {\bibfnamefont {H.}~\bibnamefont {Hafermann}},
  \bibinfo {author} {\bibfnamefont {A.~N.}\ \bibnamefont {Rubtsov}}, \bibinfo
  {author} {\bibfnamefont {M.~I.}\ \bibnamefont {Katsnelson}}, \ and\ \bibinfo
  {author} {\bibfnamefont {A.~I.}\ \bibnamefont {Lichtenstein}},\ }\href
  {\doibase 10.1103/PhysRevB.77.195105} {\bibfield  {journal} {\bibinfo
  {journal} {Phys. Rev. B}\ }\textbf {\bibinfo {volume} {77}},\ \bibinfo
  {pages} {195105} (\bibinfo {year} {2008})}\BibitemShut {NoStop}%
\bibitem [{\citenamefont {Gunnarsson}\ \emph {et~al.}(2015)\citenamefont
  {Gunnarsson}, \citenamefont {Sch\"afer}, \citenamefont {LeBlanc},
  \citenamefont {Gull}, \citenamefont {Merino}, \citenamefont {Sangiovanni},
  \citenamefont {Rohringer},\ and\ \citenamefont {Toschi}}]{Gunnarsson15}%
  \BibitemOpen
  \bibfield  {author} {\bibinfo {author} {\bibfnamefont {O.}~\bibnamefont
  {Gunnarsson}}, \bibinfo {author} {\bibfnamefont {T.}~\bibnamefont
  {Sch\"afer}}, \bibinfo {author} {\bibfnamefont {J.~P.~F.}\ \bibnamefont
  {LeBlanc}}, \bibinfo {author} {\bibfnamefont {E.}~\bibnamefont {Gull}},
  \bibinfo {author} {\bibfnamefont {J.}~\bibnamefont {Merino}}, \bibinfo
  {author} {\bibfnamefont {G.}~\bibnamefont {Sangiovanni}}, \bibinfo {author}
  {\bibfnamefont {G.}~\bibnamefont {Rohringer}}, \ and\ \bibinfo {author}
  {\bibfnamefont {A.}~\bibnamefont {Toschi}},\ }\href {\doibase
  10.1103/PhysRevLett.114.236402} {\bibfield  {journal} {\bibinfo  {journal}
  {Phys. Rev. Lett.}\ }\textbf {\bibinfo {volume} {114}},\ \bibinfo {pages}
  {236402} (\bibinfo {year} {2015})}\BibitemShut {NoStop}%
\bibitem [{\citenamefont {Wu}\ \emph {et~al.}(2017)\citenamefont {Wu},
  \citenamefont {Ferrero}, \citenamefont {Georges},\ and\ \citenamefont
  {Kozik}}]{Wu17}%
  \BibitemOpen
  \bibfield  {author} {\bibinfo {author} {\bibfnamefont {W.}~\bibnamefont
  {Wu}}, \bibinfo {author} {\bibfnamefont {M.}~\bibnamefont {Ferrero}},
  \bibinfo {author} {\bibfnamefont {A.}~\bibnamefont {Georges}}, \ and\
  \bibinfo {author} {\bibfnamefont {E.}~\bibnamefont {Kozik}},\ }\href
  {\doibase 10.1103/PhysRevB.96.041105} {\bibfield  {journal} {\bibinfo
  {journal} {Phys. Rev. B}\ }\textbf {\bibinfo {volume} {96}},\ \bibinfo
  {pages} {041105(R)} (\bibinfo {year} {2017})}\BibitemShut {NoStop}%
\bibitem [{\citenamefont {{Arzhang}}\ \emph {et~al.}(2019)\citenamefont
  {{Arzhang}}, \citenamefont {{Antipov}},\ and\ \citenamefont
  {{LeBlanc}}}]{LeBlanc19}%
  \BibitemOpen
  \bibfield  {author} {\bibinfo {author} {\bibfnamefont {B.}~\bibnamefont
  {{Arzhang}}}, \bibinfo {author} {\bibfnamefont {A.~E.}\ \bibnamefont
  {{Antipov}}}, \ and\ \bibinfo {author} {\bibfnamefont {J.~P.~F.}\
  \bibnamefont {{LeBlanc}}},\ }\href@noop {} {\enquote {\bibinfo {title}
  {{Fluctuation diagnostics of the finite temperature quasi-antiferromagnetic
  regime of the 2D Hubbard model}},}\ } (\bibinfo {year} {2019}),\ \Eprint
  {http://arxiv.org/abs/1905.07462} {arXiv:1905.07462} \BibitemShut {NoStop}%
\bibitem [{\citenamefont {Gunnarsson}\ \emph {et~al.}(2018)\citenamefont
  {Gunnarsson}, \citenamefont {Merino}, \citenamefont {Sch\"afer},
  \citenamefont {Sangiovanni}, \citenamefont {Rohringer},\ and\ \citenamefont
  {Toschi}}]{Gunnarson18}%
  \BibitemOpen
  \bibfield  {author} {\bibinfo {author} {\bibfnamefont {O.}~\bibnamefont
  {Gunnarsson}}, \bibinfo {author} {\bibfnamefont {J.}~\bibnamefont {Merino}},
  \bibinfo {author} {\bibfnamefont {T.}~\bibnamefont {Sch\"afer}}, \bibinfo
  {author} {\bibfnamefont {G.}~\bibnamefont {Sangiovanni}}, \bibinfo {author}
  {\bibfnamefont {G.}~\bibnamefont {Rohringer}}, \ and\ \bibinfo {author}
  {\bibfnamefont {A.}~\bibnamefont {Toschi}},\ }\href {\doibase
  10.1103/PhysRevB.97.125134} {\bibfield  {journal} {\bibinfo  {journal} {Phys.
  Rev. B}\ }\textbf {\bibinfo {volume} {97}},\ \bibinfo {pages} {125134}
  (\bibinfo {year} {2018})}\BibitemShut {NoStop}%
\bibitem [{\citenamefont {Abbamonte}\ \emph {et~al.}(2012)\citenamefont
  {Abbamonte}, \citenamefont {Demler}, \citenamefont {Davis},\ and\
  \citenamefont {Campuzano}}]{Abbamonte12_stripe}%
  \BibitemOpen
  \bibfield  {author} {\bibinfo {author} {\bibfnamefont {P.}~\bibnamefont
  {Abbamonte}}, \bibinfo {author} {\bibfnamefont {E.}~\bibnamefont {Demler}},
  \bibinfo {author} {\bibfnamefont {J.~S.}\ \bibnamefont {Davis}}, \ and\
  \bibinfo {author} {\bibfnamefont {J.-C.}\ \bibnamefont {Campuzano}},\ }\href
  {\doibase https://doi.org/10.1016/j.physc.2012.04.006} {\bibfield  {journal}
  {\bibinfo  {journal} {Physica C: Superconductivity}\ }\textbf {\bibinfo
  {volume} {481}},\ \bibinfo {pages} {15 } (\bibinfo {year} {2012})},\ \bibinfo
  {note} {stripes and Electronic Liquid Crystals in Strongly Correlated
  Materials}\BibitemShut {NoStop}%
\bibitem [{\citenamefont {Kohsaka}\ \emph {et~al.}(2007)\citenamefont
  {Kohsaka}, \citenamefont {Taylor}, \citenamefont {Fujita}, \citenamefont
  {Schmidt}, \citenamefont {Lupien}, \citenamefont {Hanaguri}, \citenamefont
  {Azuma}, \citenamefont {Takano}, \citenamefont {Eisaki}, \citenamefont
  {Takagi}, \citenamefont {Uchida},\ and\ \citenamefont
  {Davis}}]{Kohsaka07_stripe}%
  \BibitemOpen
  \bibfield  {author} {\bibinfo {author} {\bibfnamefont {Y.}~\bibnamefont
  {Kohsaka}}, \bibinfo {author} {\bibfnamefont {C.}~\bibnamefont {Taylor}},
  \bibinfo {author} {\bibfnamefont {K.}~\bibnamefont {Fujita}}, \bibinfo
  {author} {\bibfnamefont {A.}~\bibnamefont {Schmidt}}, \bibinfo {author}
  {\bibfnamefont {C.}~\bibnamefont {Lupien}}, \bibinfo {author} {\bibfnamefont
  {T.}~\bibnamefont {Hanaguri}}, \bibinfo {author} {\bibfnamefont
  {M.}~\bibnamefont {Azuma}}, \bibinfo {author} {\bibfnamefont
  {M.}~\bibnamefont {Takano}}, \bibinfo {author} {\bibfnamefont
  {H.}~\bibnamefont {Eisaki}}, \bibinfo {author} {\bibfnamefont
  {H.}~\bibnamefont {Takagi}}, \bibinfo {author} {\bibfnamefont
  {S.}~\bibnamefont {Uchida}}, \ and\ \bibinfo {author} {\bibfnamefont {J.~C.}\
  \bibnamefont {Davis}},\ }\href {\doibase 10.1126/science.1138584} {\bibfield
  {journal} {\bibinfo  {journal} {Science}\ }\textbf {\bibinfo {volume}
  {315}},\ \bibinfo {pages} {1380} (\bibinfo {year} {2007})}\BibitemShut
  {NoStop}%
\bibitem [{\citenamefont {Fujita}\ \emph {et~al.}(2004)\citenamefont {Fujita},
  \citenamefont {Goka}, \citenamefont {Yamada}, \citenamefont {Tranquada},\
  and\ \citenamefont {Regnault}}]{Fujita04_stripe}%
  \BibitemOpen
  \bibfield  {author} {\bibinfo {author} {\bibfnamefont {M.}~\bibnamefont
  {Fujita}}, \bibinfo {author} {\bibfnamefont {H.}~\bibnamefont {Goka}},
  \bibinfo {author} {\bibfnamefont {K.}~\bibnamefont {Yamada}}, \bibinfo
  {author} {\bibfnamefont {J.~M.}\ \bibnamefont {Tranquada}}, \ and\ \bibinfo
  {author} {\bibfnamefont {L.~P.}\ \bibnamefont {Regnault}},\ }\href {\doibase
  10.1103/PhysRevB.70.104517} {\bibfield  {journal} {\bibinfo  {journal} {Phys.
  Rev. B}\ }\textbf {\bibinfo {volume} {70}},\ \bibinfo {pages} {104517}
  (\bibinfo {year} {2004})}\BibitemShut {NoStop}%
\bibitem [{\citenamefont {Parker}\ \emph {et~al.}(2010)\citenamefont {Parker},
  \citenamefont {Aynajian}, \citenamefont {da~Silva~Neto}, \citenamefont
  {Pushp}, \citenamefont {Ono}, \citenamefont {Wen}, \citenamefont {Xu},
  \citenamefont {Gu},\ and\ \citenamefont {Yazdani}}]{Parker10}%
  \BibitemOpen
  \bibfield  {author} {\bibinfo {author} {\bibfnamefont {C.~V.}\ \bibnamefont
  {Parker}}, \bibinfo {author} {\bibfnamefont {P.}~\bibnamefont {Aynajian}},
  \bibinfo {author} {\bibfnamefont {E.~H.}\ \bibnamefont {da~Silva~Neto}},
  \bibinfo {author} {\bibfnamefont {A.}~\bibnamefont {Pushp}}, \bibinfo
  {author} {\bibfnamefont {S.}~\bibnamefont {Ono}}, \bibinfo {author}
  {\bibfnamefont {J.}~\bibnamefont {Wen}}, \bibinfo {author} {\bibfnamefont
  {Z.}~\bibnamefont {Xu}}, \bibinfo {author} {\bibfnamefont {G.}~\bibnamefont
  {Gu}}, \ and\ \bibinfo {author} {\bibfnamefont {A.}~\bibnamefont {Yazdani}},\
  }\href {\doibase 10.1038/nature09597} {\bibfield  {journal} {\bibinfo
  {journal} {Nature}\ }\textbf {\bibinfo {volume} {468}},\ \bibinfo {pages}
  {677} (\bibinfo {year} {2010})}\BibitemShut {NoStop}%
\bibitem [{\citenamefont {Lawler}\ \emph {et~al.}(2010)\citenamefont {Lawler},
  \citenamefont {Fujita}, \citenamefont {Lee}, \citenamefont {Schmidt},
  \citenamefont {Kohsaka}, \citenamefont {Kim}, \citenamefont {Eisaki},
  \citenamefont {Uchida}, \citenamefont {Davis}, \citenamefont {Sethna},\ and\
  \citenamefont {Kim}}]{Lawler10_stripe}%
  \BibitemOpen
  \bibfield  {author} {\bibinfo {author} {\bibfnamefont {M.~J.}\ \bibnamefont
  {Lawler}}, \bibinfo {author} {\bibfnamefont {K.}~\bibnamefont {Fujita}},
  \bibinfo {author} {\bibfnamefont {J.}~\bibnamefont {Lee}}, \bibinfo {author}
  {\bibfnamefont {A.~R.}\ \bibnamefont {Schmidt}}, \bibinfo {author}
  {\bibfnamefont {Y.}~\bibnamefont {Kohsaka}}, \bibinfo {author} {\bibfnamefont
  {C.~K.}\ \bibnamefont {Kim}}, \bibinfo {author} {\bibfnamefont
  {H.}~\bibnamefont {Eisaki}}, \bibinfo {author} {\bibfnamefont
  {S.}~\bibnamefont {Uchida}}, \bibinfo {author} {\bibfnamefont {J.~C.}\
  \bibnamefont {Davis}}, \bibinfo {author} {\bibfnamefont {J.~P.}\ \bibnamefont
  {Sethna}}, \ and\ \bibinfo {author} {\bibfnamefont {E.-A.}\ \bibnamefont
  {Kim}},\ }\href {https://doi.org/10.1038/nature09169} {\bibfield  {journal}
  {\bibinfo  {journal} {Nature}\ }\textbf {\bibinfo {volume} {466}},\ \bibinfo
  {pages} {347 EP } (\bibinfo {year} {2010})}\BibitemShut {NoStop}%
\bibitem [{\citenamefont {Hashimoto}\ \emph {et~al.}(2010)\citenamefont
  {Hashimoto}, \citenamefont {He}, \citenamefont {Tanaka}, \citenamefont
  {Testaud}, \citenamefont {Meevasana}, \citenamefont {Moore}, \citenamefont
  {Lu}, \citenamefont {Yao}, \citenamefont {Yoshida}, \citenamefont {Eisaki},
  \citenamefont {Devereaux}, \citenamefont {Hussain},\ and\ \citenamefont
  {Shen}}]{Hashimoto10_stripe}%
  \BibitemOpen
  \bibfield  {author} {\bibinfo {author} {\bibfnamefont {M.}~\bibnamefont
  {Hashimoto}}, \bibinfo {author} {\bibfnamefont {R.-H.}\ \bibnamefont {He}},
  \bibinfo {author} {\bibfnamefont {K.}~\bibnamefont {Tanaka}}, \bibinfo
  {author} {\bibfnamefont {J.-P.}\ \bibnamefont {Testaud}}, \bibinfo {author}
  {\bibfnamefont {W.}~\bibnamefont {Meevasana}}, \bibinfo {author}
  {\bibfnamefont {R.~G.}\ \bibnamefont {Moore}}, \bibinfo {author}
  {\bibfnamefont {D.}~\bibnamefont {Lu}}, \bibinfo {author} {\bibfnamefont
  {H.}~\bibnamefont {Yao}}, \bibinfo {author} {\bibfnamefont {Y.}~\bibnamefont
  {Yoshida}}, \bibinfo {author} {\bibfnamefont {H.}~\bibnamefont {Eisaki}},
  \bibinfo {author} {\bibfnamefont {T.~P.}\ \bibnamefont {Devereaux}}, \bibinfo
  {author} {\bibfnamefont {Z.}~\bibnamefont {Hussain}}, \ and\ \bibinfo
  {author} {\bibfnamefont {Z.-X.}\ \bibnamefont {Shen}},\ }\href
  {https://doi.org/10.1038/nphys1632} {\bibfield  {journal} {\bibinfo
  {journal} {Nature Physics}\ }\textbf {\bibinfo {volume} {6}},\ \bibinfo
  {pages} {414 EP } (\bibinfo {year} {2010})}\BibitemShut {NoStop}%
\bibitem [{\citenamefont {{Daou}}\ \emph {et~al.}(2010)\citenamefont {{Daou}},
  \citenamefont {{Chang}}, \citenamefont {{Leboeuf}}, \citenamefont
  {{Cyr-Choini{\`e}re}}, \citenamefont {{Lalibert{\'e}}}, \citenamefont
  {{Doiron-Leyraud}}, \citenamefont {{Ramshaw}}, \citenamefont {{Liang}},
  \citenamefont {{Bonn}},\ and\ \citenamefont {{Hardy}}}]{Daou10_stripe}%
  \BibitemOpen
  \bibfield  {author} {\bibinfo {author} {\bibfnamefont {R.}~\bibnamefont
  {{Daou}}}, \bibinfo {author} {\bibfnamefont {J.}~\bibnamefont {{Chang}}},
  \bibinfo {author} {\bibfnamefont {D.}~\bibnamefont {{Leboeuf}}}, \bibinfo
  {author} {\bibfnamefont {O.}~\bibnamefont {{Cyr-Choini{\`e}re}}}, \bibinfo
  {author} {\bibfnamefont {F.}~\bibnamefont {{Lalibert{\'e}}}}, \bibinfo
  {author} {\bibfnamefont {N.}~\bibnamefont {{Doiron-Leyraud}}}, \bibinfo
  {author} {\bibfnamefont {B.~J.}\ \bibnamefont {{Ramshaw}}}, \bibinfo {author}
  {\bibfnamefont {R.}~\bibnamefont {{Liang}}}, \bibinfo {author} {\bibfnamefont
  {D.~A.}\ \bibnamefont {{Bonn}}}, \ and\ \bibinfo {author} {\bibfnamefont
  {W.~N.}\ \bibnamefont {{Hardy}}},\ }\href {\doibase 10.1038/nature08716}
  {\bibfield  {journal} {\bibinfo  {journal} {\nat}\ }\textbf {\bibinfo
  {volume} {463}},\ \bibinfo {pages} {519} (\bibinfo {year} {2010})},\ \Eprint
  {http://arxiv.org/abs/0909.4430} {0909.4430} \BibitemShut {NoStop}%
\end{thebibliography}%
\end{document}